\documentclass[12pt]{article}

\def\gappeq{\mathrel{\rlap {\raise.5ex\hbox{$>$}}
{\lower.5ex\hbox{$\sim$}}}}

\def\lappeq{\mathrel{\rlap{\raise.5ex\hbox{$<$}}
{\lower.5ex\hbox{$\sim$}}}}

\begin{document}
\setcounter{footnote}{0}
\newcommand{\mycomm}[1]{\hfill\break
 $\phantom{a}$\kern-3.5em{\tt===$>$ \bf #1}\hfill\break}
\newcommand{\mycommA}[1]{\hfill\break
$\phantom{a}$\kern-3.5em{\tt***$>$ \bf #1}\hfill\break}
\renewcommand{\thefootnote}{\fnsymbol{footnote}}

\catcode`\@=11 
\def\lsim{\mathrel{\mathpalette\@versim<}}
\def\gsim{\mathrel{\mathpalette\@versim>}}
\def\@versim#1#2{\vcenter{\offinterlineskip
        \ialign{$\m@th#1\hfil##\hfil$\crcr#2\crcr\sim\crcr } }}
\catcode`\@=12 
\def\beq{\begin{equation}}
\def\eeq{\end{equation}}
\def\MSbar {\hbox{$\overline{\hbox{\tiny MS}}\,$}}
\def\eff{\hbox{\tiny eff}}
\def\FP{\hbox{\tiny FP}}
\def\PV{\hbox{\tiny PV}}
\def\IR{\hbox{\tiny IR}}
\def\UV{\hbox{\tiny UV}}
\def\ECH{\hbox{\tiny ECH}}
\def\APT{\hbox{\tiny APT}}
\def\QCD{\hbox{\tiny QCD}}
\def\CMW{\hbox{\tiny CMW}}
\def\pinch{\hbox{\tiny pinch}}
\def\brem{\hbox{\tiny brem}}
\def\V{\hbox{\tiny V}}
\def\BLM{\hbox{\tiny BLM}}
\def\NLO{\hbox{\tiny NLO}}
\def\res{\hbox{\tiny res}}
\def\PT{\hbox{\tiny PT}}
\def\PA{\hbox{\tiny PA}}
\def\1loop{\hbox{\tiny 1-loop}}
\def\2loop{\hbox{\tiny 2-loop}}
\def\mysim{\kern -.1667em\lower0.8ex\hbox{$\tilde{\phantom{a}}$}}
\def\a{\bar{a}}

\begin{titlepage}
\begin{flushright}
{\small S 746-1099}\\ 
{\small LPT-Orsay 00-09}\\
{\small SLAC-PUB-8362}\\ 
{\small CERN-TH/2000-032}\\
{\small October 2000}

\end{flushright}
\vspace{.13in}

\begin{centering} 
{\Large {\bf Disentangling running coupling \\
and conformal effects in QCD\footnote{Research
supported in part by the EC
program ``Training and Mobility of Researchers'', Network
``QCD and Particle Structure'', contract ERBFMRXCT980194
and the U.S. Department of Energy,
contract DE--AC03--76SF00515.} }} \\
\end{centering}
\vspace{.39in}

{\bf S.J. Brodsky}$^{(1)}$, \,\,
\,{\bf E. Gardi}$^{(2,3,4)}\footnote{present address: (4).}$,
\,\,\,{\bf G. Grunberg}$^{(2)}$\,\,\,and\,\,\,{\bf J. Rathsman}$^{(4)}$

\vspace{.43in}
{\small 
\noindent
$^{(1)}$ Stanford Linear Accelerator Center, Stanford University, Stanford, 
CA 94309, USA

\vspace{.20in}
\noindent
$^{(2)}$
 Centre de Physique Th\'eorique de l'Ecole Polytechnique\footnote{CNRS
 UMR C7644}, 91128 Palaiseau Cedex, France

\vspace{.20in}
\noindent
$^{(3)}$
Laboratoire de Physique Th\'eorique\footnote{CNRS UMR 8627},
Universit\'e de Paris XI, 91405 Orsay Cedex, France

\vspace{0.20in} 
\noindent
$^{(4)}$ TH Division, CERN, CH-1211 Geneva 23, Switzerland
}

\vspace{.24in}

\noindent
{\small {\bf Abstract:\,}
We investigate the relation between a postulated skeleton expansion 
and the conformal limit of QCD. We begin by developing some consequences 
of an Abelian-like skeleton expansion, which allows one to 
disentangle running-coupling effects from the remaining skeleton 
coefficients. The latter are by construction renormalon-free, and hence 
hopefully better behaved.  We consider a simple ansatz for the expansion, 
where an observable is written as a sum of integrals over the running-coupling.
We show that in this framework one
can set a unique Brodsky-Lepage-Mackenzie (BLM) scale-setting procedure as
an approximation to the running-coupling integrals, where the BLM
coefficients coincide with the skeleton ones. 
Alternatively, the running-coupling integrals can be approximated 
using the effective charge method. We discuss the limitations in 
disentangling running coupling effects in the absence of a diagrammatic
construction of the skeleton expansion. Independently of the assumed skeleton 
structure we show that BLM coefficients coincide with the conformal
coefficients defined in the small $\beta_0$ (Banks-Zaks) limit 
where a perturbative infrared fixed-point is present. This interpretation 
of the BLM coefficients should explain their previously observed 
simplicity and smallness. Numerical examples are critically discussed.
\vspace{.05in}}
\end{titlepage}
\newpage

\section {Introduction}

The large-order behavior of a perturbative expansion in gauge theories is
inevitably dominated by the factorial growth of renormalon
diagrams~\cite{tHooft,Mueller,Zakharov,Beneke}. In the case of quantum
chromodynamics (QCD), the coefficients of perturbative expansions in the QCD
coupling $\alpha_s$ can increase dramatically even at low orders. This
fact, together with the apparent freedom in the choice of
renormalization scheme and renormalization scales, limits the
predictive power of perturbative calculations, even in
applications involving large momentum transfer where $\alpha_s$ is
effectively small.

A number of theoretical approaches have been developed to
reorganize the perturbative expansions in an effort to improve
the predictability of perturbative QCD. For example,
optimized scale and scheme choices have been proposed, such as the
method of effective charges [ECH]~\cite{ECH},
the principle of minimal sensitivity [PMS]~\cite{PMS},
and the Brodsky-Lepage-Mackenzie [BLM] scale-setting
prescription~\cite{BLM} and its
generalizations~\cite{GruKat}--\cite{Brodsky:1998mf}.
More recent developments \cite{Beneke} include the resummation of the
formally divergent renormalon series and the parameterization of related
higher-twist power-suppressed contributions.

In general, a factorially divergent renormalon series arises when one
integrates over the logarithmically running coupling
$\alpha_s(k^2)$ in a loop diagram. Such contributions do not
occur in conformally invariant theories which have a constant coupling.
Of course, in the physical theory, the QCD coupling does run.
Nevertheless, relying on a postulated ``dressed skeleton expansion'',
we shall show that a conformal series is directly relevant to physical
QCD predictions.

In quantum electrodynamics the dressed skeleton expansion can replace
the standard perturbative expansion. The skeleton diagrams are defined
as those Feynman graphs where the three-point vertex and the lepton
and photon propagators have no substructure \cite{Bjorken_Drell}.
Thanks to the QED Ward identity, the renormalization of the vertex cancels
against the lepton self-energy, while the effect of dressing the
photons in the skeleton diagrams by vacuum polarization insertions
can be computed by integrating over the Gell-Mann Low effective
charge $\bar{\alpha}(k^2)$. The perturbative coefficients
defined from the skeleton graphs themselves are conformal -- they
correspond to the series in a theory with a zero $\beta$ function.
Therefore they are entirely free of running coupling effects such as
renormalons. Each term in the dressed skeleton expansion resums
renormalon diagrams to all orders in a renormalization scheme
invariant way. The resummation ambiguity, which is associated with
scales where the coupling becomes strong, can be resolved only at the
non-perturbative level.

In QCD, a skeleton expansion can presumably be constructed based on
several different dressed Green functions (see \cite{Baker_Lee}).
A much more interesting possibility, which is yet speculative, is the 
existence of an Abelian-like skeleton expansion, with only one effective 
charge function. The construction of such an expansion is not
straightforward due to the presence of gluon self-interaction diagrams
and the essential difference between vacuum polarization insertions and
charge renormalization. 
Nevertheless, at the one-loop level there is a diagrammatic algorithm, 
the so-called ``pinch technique''~\cite{pinch}, that allows one to 
identify in every non-Abelian diagram the part which can
be absorbed into the renormalization of the effective gluon propagator.
The sum of all the vacuum-polarization-like parts turns out to be
gauge-invariant, thus defining a natural candidate for the non-Abelian
equivalent of the Gell-Mann Low effective charge,
$\bar{\alpha}_s(k^2)$. Moreover, the
pinch technique leads to Ward identities similar to those of the Abelian 
theory: after the vacuum-polarization-like parts have been taken
into account, the remaining vertex correction cancels against the 
quark self-energy. 
In this way the pinch technique achieves the first step in the construction of 
an Abelian-like skeleton expansion. Recently there have been some
encouraging developments \cite{Watson,papa_2_loops} 
in the application of the pinch technique beyond one-loop and its
possible relation to the background field method. The hope is that 
these techniques will eventually provide a proof of existence for the
skeleton expansion as well as an all-order constructive definition  
for the non-Abelian skeleton structure and the non-Abelian
skeleton effective charge $\bar{\alpha}_s(k^2)$. 

In this paper, we shall postulate that an Abelian-like
skeleton expansion can be defined at arbitrary order in QCD.
We shall not deal here with the diagrammatic construction of the skeleton
expansion but rather restrict ourselves to the consequences which 
follow from such a structure. To this end we will introduce a simple
ansatz for the skeleton expansion, where similarly to the Abelian case,
a generic observable is written as a sum of integrals over the
running coupling. As in QED, we can then identify running 
coupling effects to all orders, and treat 
them separately from the conformal part of the perturbative expansion.
A considerable simplification is achieved, for instance, by assuming
that the dependence on the number of light quark flavors $N_f$ originates only
in the running coupling itself, as in the Abelian theory with
light-by-light diagrams being excluded. As a consequence, the 
coefficients appearing in the assumed skeleton expansion are $N_f$
independent. By construction these skeleton coefficients are {\em free of
renormalons} and are therefore expected to be better behaved.  
We will show that they have a simple interpretation in the presence of 
a perturbative infrared fixed-point, as occurs in the small $\beta_0$ 
limit: they are the ``conformal'' coefficients in the series
relating the fixed-point value of the observable under consideration
with that of the skeleton effective charge. Thus, given the assumption
that these coefficients are $N_f$ independent, they can be obtained 
from the standard perturbative coefficients using the Banks-Zaks 
expansion~\cite{BZ,BZ_grunberg,White}, where the fixed-point coupling 
is expanded in powers of $\beta_0$.

The conformal series can be seen as a template~\cite{Brodsky:1999gm,Lu}
for {\em physical} QCD predictions, where instead of the fixed coupling
one has at each order a weighted average of the skeleton effective charge
$\bar{\alpha}_s(k^2)$ with respect to an observable- (and order-)
dependent momentum distribution function.
The momentum integral corresponding to each skeleton term is
renormalization-scheme invariant. 
Had the skeleton effective charge been known at all scales, this integral 
could have been unambiguously evaluated, thus including 
both perturbative and non-perturbative contributions. 
In practice it can be evaluated up to
power-suppressed ambiguities, which originate in the infrared where the 
coupling becomes strong. These infrared renormalon ambiguities can 
be resolved only by explicitly taking non-perturbative effects into account.
Since such effects cannot be calculated with present methods, they can only 
be parametrized. Indeed, a natural parametrization in the form of an infrared 
finite coupling~\cite{DMW} emerges from the structure of the skeleton 
integral. This way the skeleton expansion gives a natural framework in which
renormalon resummation and the analysis of non-perturbative power
corrections are performed together~\cite{Grunberg-pow,thrust}.

As an alternative to computing a dressed skeleton integral, one can
approximate it by evaluating the coupling at the BLM scale~\cite{BLM}, 
in analogy to the mean-value theorem~\cite{Lepage:1993xa}. 
By going to higher orders in the perturbative expansion, this 
approximation can be systematically improved. Another possibility
considered here is to approximate each dressed skeleton term separately 
using the effective charge approach. This approach is tailored
\cite{ECH} to deal with 
running coupling effects and it by-passes remaining scheme and scale 
setting ambiguities in the power series expressions for the BLM scales.
Assuming a simple form of the skeleton expansion, running coupling
effects can be disentangled from the remaining conformal expansion by
tracing the $N_f$ dependence of the coefficients. In this case
BLM scale-setting (or the ECH alternative) can be applied to 
a generic QCD observable based on the knowledge of the first few 
coefficients. In the general case disentangling running 
coupling effects becomes more involved, and it eventually requires 
a diagrammatic construction of the skeleton expansion. 
We emphasize that both the BLM scale-setting method and the suggested 
ECH method remain on the perturbative level and, as opposed to the 
infrared finite coupling approach mentioned above, these methods are not 
particularly suited to deal with renormalon ambiguities and the related 
power-corrections. 

BLM scale-setting can also be applied to the perturbative relation 
between the effective charges of two physical observables. This results in a
specific ``commensurate scale relation''~\cite{CSR} between the two
quantities. The coefficients appearing in such relations are conformal
and, as guaranteed by the
transitivity property of the renormalization group, they do not 
depend on the intermediate scheme used.
This way conformal relations appear to be relevant for real-world QCD
predictions even in the absence of complete understanding of the
underlying skeleton structure. 
In the case of the Crewther relation~\cite{Crewther,Crewther_obs,BLM_Crewther},
which connects the effective charges of the $e^+ e^-$ annihilation cross 
section to the Bjorken and Gross-Llewellyn Smith sum rules for deep inelastic
scattering, the conformal relation is simply a geometric series. This
example highlights the power of characterizing QCD perturbative
expansions in terms of conformal coefficients.

This paper is organized as follows: we begin in section~2 by recalling 
the concept of the skeleton expansion in the Abelian case \cite{Bjorken_Drell}
and stating the main assumptions concerning the non-Abelian case.
We continue, in section~3, by reviewing the standard BLM scale-setting
procedure and recalling the ambiguity of the procedure beyond the
next-to-leading order. We then show how this ambiguity is resolved
upon assuming a skeleton expansion, provided we work in the appropriate
renormalization scheme, the ``skeleton scheme'', and require a one-to-one
correspondence between the terms in the BLM series and the dressed skeletons.
We also discuss in this section the limitations of applying the formal 
BLM procedure in the absence of a diagrammatic construction of the 
skeleton expansion. 

In section~4 we present an alternative to 
performing an explicit scale setting, by using the ECH method as a tool
to  resum running-coupling effects within the framework of the 
assumed skeleton expansion. Close  connections between the two 
approaches are pointed out. In Appendix A we look
at the original ECH approach from the point of view of the
skeleton expansion, comparing it to the application of the ECH method to the 
leading skeleton term. We also calculate there the $\beta_0=0$ limit of 
the skeleton coupling $\beta$ function coefficient $\bar{\beta}_2$.

We then come to the main subject of the paper, the relation between 
the coefficients which remain after applying BLM 
scale-setting and the conformal limit of QCD.
We derive (section~5) a relation between these BLM
coefficients and the conformal coefficients defined in the
infrared limit in the conformal window, where a non-trivial perturbative
fixed-point exists~\cite{Gross:1973ju}--\cite{super}.
In section~6 we show explicitly that the conformal coefficients, 
calculated using the Banks-Zaks expansion, are the same as the ones in 
the BLM series. In section~7 we recall previous observations
concerning the smallness of conformal and Banks-Zaks coefficients,
and examine whether this apparent convergence can be explained by the 
absence of renormalons in such relations. 
The conclusions are given in section~8.

\section{Renormalons and the skeleton expansion}

Consider a Euclidean QED observable $a_R(Q^2)$, which depends on a
single external space-like momentum $Q^2$ and is normalized as an
effective charge. The perturbative expansion in a generic
renormalization scheme is then given by,
\beq
a_R(Q^2) = a(\mu^2) +r_1 {a(\mu^2)}^2+r_2 {a(\mu^2)}^3+\cdots ,
\label{standard_exp}
\eeq
where $a=\alpha/\pi$ and $\mu$ is the renormalization scale.

The perturbative series can be reorganized and written in the form of
a skeleton expansion
\beq
a_R(Q^2) = R_0(Q^2) + s_1 R_1(Q^2) + s_2 R_2(Q^2)+\cdots,
\label{skeleton_exp}
\eeq
where the first term, $R_0$, corresponds to a single dressed photon:
it is the infinite set of ``renormalon diagrams'' obtained by all
possible vacuum polarization insertions into a single photon
line. The second term, $s_1R_1$,
corresponds to a double dressed-photon exchange and so on.
In QED, vacuum polarization insertions amount to charge renormalization.
Thus $R_0$ can be written as
\beq
R_0(Q^2)\equiv \int_0^{\infty} \a (k^2) \,
\phi_0\left({k^2}/{Q^2}\right) \frac{dk^2}{k^2}
\label{R_0}
\eeq
where $k^2$ is the virtuality of the exchanged photon,
$\a(k^2)$ is the Gell-Mann Low effective charge representing the full
propagator, and $\phi_0$ is the
(observable dependent) Feynman integrand for a single photon exchange diagram,
which is interpreted as the photon momentum distribution
function~\cite{Neu}. Similarly, $R_1$ is given by
\begin{equation}
R_1(Q^2)\equiv
\int_{0}^\infty
\bar{a}(k_1^2)\,\bar{a}(k_2^2)\,
\phi_1\left(k_1^2/Q^2,k_2^2/Q^2\right)
{dk_1^2\over k_1^2}\,{dk_2^2\over k_2^2}
\label{R_1}
\end{equation}
and so on.

For convenience the normalization of $\phi_i$ in $R_i(Q^2)$
has been set to 1 such that the $R_i(Q^2)$ in (\ref{skeleton_exp})
have an expansion $R_i(Q^2)=\bar{a}(Q^2)^{i+1}+\cdots$.
For example, the normalization of $\phi_0(k^2/Q^2)$ in $R_0$ is
\beq
\int_0^{\infty} \phi_0\left({k^2}/{Q^2}\right) \frac{dk^2}{k^2}=1.
\label{norm_phi_0}
\eeq

In QED fermion loops appear either dressing the exchanged photons or
in light-by-light type diagrams, where they are attached to four or more
photons (an even number). Barring the latter, the
dependence on the number of massless fermion flavors $N_f$ is fully
contained in the Gell-Mann Low effective charge. It follows that the skeleton
coefficients $s_i$ as well as the momentum distribution functions
$\phi_i$ are entirely free of $N_f$ dependence. Light-by-light type
diagrams have to be treated separately, as the starting point of new
skeleton structures.

\setcounter{footnote}{0}
The skeleton expansion (\ref{skeleton_exp}) is a renormalization
group invariant expansion: each term is by itself scheme
invariant. This is in contrast with the standard scale and scheme
dependent perturbative expansion (\ref{standard_exp}).
The renormalons in (\ref{standard_exp}) can be obtained upon
expansion of the dressed skeleton terms in (\ref{skeleton_exp}) in
some scheme. Let us consider first the leading skeleton (\ref{R_0})
and examine, for simplicity, its expansion in $\a(Q^2)$.
We assume that $\a(k^2)$ obeys the renormalization group equation,
\beq
\frac{d\a(k^2)}{d\ln k^2}=-\left(\beta_0\a(k^2)^2+\beta_1\a(k^2)^3
+\bar{\beta}_2\a(k^2)^4+\cdots\right)\,\equiv\,\bar{\beta}(\a)
\label{beta_func}
\eeq
where $\beta_0$ is negative in QED and positive in QCD.
Then $\a(k^2)$ can be expanded as
\beq
\a(k^2)=\a(Q^2)+\beta_0t\a(Q^2)^2+\left(\beta_1t+\beta_0^2t^2\right)\a(Q^2)^3
+\left(\bar{\beta}_2t+\frac52\beta_1\beta_0t^2+\beta_0^3t^3\right)
\a(Q^2)^4+\cdots
\label{rescale}
\eeq
where $t\equiv - \ln(k^2/Q^2)$.
Inserting this in eq.~(\ref{R_0}) under the integration sign we obtain
\begin{eqnarray}
\label{R_0_expand}
R_0(Q^2)=\a(Q^2)&+& r_1^{(1)}
\beta_0\a(Q^2)^2+\left(r_2^{(2)}\beta_0^2+ r_1^{(1)}\beta_1\right)
\a(Q^2)^3\nonumber \\  &+&
\left(r_3^{(3)}\beta_0^3+\frac52r_2^{(2)}\beta_1\beta_0+r_1^{(1)}\bar{\beta}_2
\right)\a(Q^2)^4
+\cdots
\end{eqnarray}
where
\beq
r_i^{(i)}\equiv \int_0^{\infty}
\left[-\ln\left(k^2/Q^2\right)\right]^i\,\phi_0(k^2/Q^2)\,\frac{dk^2}{k^2} .
\label{r_i^i}
\eeq

We note that in the large $N_f$ (large $\beta_0$) limit\footnote{In QCD,
Abelian correspondence in the large $N_f$ limit
requires that the coefficient $\bar{\beta}_i$ of the skeleton coupling
$\beta$ function (\ref{beta_func}) would not
contain $N_f^{i+1}$.  It has to be a polynomial of order $N_f^{i}$ in $N_f$.
This would guarantee that in the large $N_f$ limit $\bar{\beta}(\bar{a})$ is
just the one-loop $\beta$ function.  Note that while some schemes (e.g.
$\overline{\rm MS}$ and static potential effective charge) have this
property, generic effective charges (defined through observable
quantities) do not.
This property of the skeleton scheme is used making the identification
of $r_i^{(i)}$ in (\ref{R_0_expand}) as the large $N_f$ coefficients.},
the perturbative coefficients $r_i=r_i^{(i)}$ and thus\footnote{We
comment that the sub-leading terms in $1/N_f$ in (\ref{R_0_expand})
of the form $\beta_1\beta_0^{i-2}$ were computed to all-orders in
\cite{BBB}. However, other terms which involve higher order
coefficients of the $\beta$ function contribute at the same level in $1/N_f$.
}
\beq
\left.a_R(Q^2)\right\vert_{{\rm large} \,\,\beta_0}
=\a(Q^2)\,\left[\sum_{i=0}^{\infty}r_i^{(i)}
\left( \beta_0\a(Q^2)\right)^i+
{\cal O}\left(1/\beta_0\right)\right].
\eeq
At large orders $i\gg 1$,
both small and large momentum regions become dominant
in (\ref{r_i^i}), giving rise to the characteristic renormalon
factorial divergence ($r_i^{(i)}\sim i!$).
As mentioned above, this is believed to be the dominant source of
divergence of the perturbative expansion (\ref{standard_exp}).
On the other hand, in the skeleton expansion (\ref{skeleton_exp})
the renormalons are by definition resummed and so the remaining coefficients
$s_i$ should be free of this divergence. These coefficients are
expected to increase much slower leading to a better behaved expansion.

As mentioned in the introduction, the generalization of the Abelian
skeleton expansion to QCD is not straightforward.
Diagrammatically, the skeleton expansion in QCD has a simple realization only
in the large $N_f$ limit where gluon self-interaction contributions
are negligible so that the theory resembles QED\footnote{This can also be
understood from the $C_A \to 0$ limit discussed in
ref.~\cite{Brodsky:1997jk}.}.  In the framework of renormalon calculus,
one returns from the large $N_f$ limit to real world QCD by replacing
$N_f$ with the linear combination of $N_f$ and $C_A=N_c$ which appears in the
leading coefficient \cite{oneloop} of the $\beta$ function,
\beq
\beta_0=\frac14\left (\frac{11}{3}C_A-\frac23N_f\right).
\label{beta_0}
\eeq
This replacement, usually called ``naive non-Abelianization''
\cite{Broadhurst,Neu,BBB,LTM},
amounts to taking into account a gauge invariant set of diagrams
which is responsible for the one-loop running of the coupling constant.

To go beyond the ``naive non-Abelianization'' level constructing an
Abelian-like skeleton expansion in QCD, one needs a method
to identify skeleton structures and to isolate 
vacuum-polarization-like insertions which are responsible for
the running of the coupling at any order.  The pinch technique
\cite{pinch,Watson,papa_2_loops} may provide a systematic way to make
this identification.
The resulting set of skeleton structures would surely be larger than in the
Abelian theory.  It may include, for example, fermion loops attached to
an odd number of gluons, which vanish in the Abelian limit.  Like
Abelian light-by-light type diagrams, these structures should be treated
separately.  As opposed to the Abelian theory, where light-by-light type
diagrams are distinguished by their characteristic dependence on the charges,
in the non-Abelian case these structures may not be separable based only on
their color group structure. 
We assume that there is a unique (gauge invariant) way to identify 
skeleton structures in QCD, making eq.~(\ref{skeleton_exp}) relevant. 
In much of the discussion that follows we shall further 
make the assumption that the entire dependence of $a_R(Q^2)$ on $N_f$ is 
through the running coupling. Thus in our ansatz 
$s_i$ and $\phi_i$ are $N_f$ independent, just like in the Abelian 
case with light-by-light diagrams being excluded. Of course, the
class of diagrams containing fermion loops as part 
of the skeleton structure should eventually be taken into account.

Another simplifying assumption we made already in the Abelian case is
that our ansatz~(\ref{skeleton_exp}) contains only one skeleton 
at each order, whereas in general there will be several skeletons 
contributing at each order. The simplest example is $e^-e^-$ scattering 
with both $t$- and $u$-channel exchange. Several skeletons at the same order 
also occur in single-scale observables considered here, and therefore  
(\ref{skeleton_exp}) should be generalized accordingly. We shall return
to this point in the next section.

We stress that the coupling constant $\a(k^2)$ in (\ref{R_0}) is
understood to be a {\em specific} effective charge, in analogy to the
Gell-Mann Low effective charge in QED. This ``skeleton effective
charge'' $\a(k^2)$ should be defined diagrammatically order by order
in perturbation theory.  In the framework of the pinch technique,
$\a(k^2)$ has been identified at the
one-loop level\footnote{This means that the
corresponding QCD scale $\bar{\Lambda}$ is identified.},
e.g.  it is related to the $\overline{\rm MS}$ coupling by
\begin{equation}
\bar{a}(k^2)= a_{\MSbar}(\mu^2) \,+\,
\left[-\beta_0 \left(\log{k^2\over
\mu^2}-\frac53\right)+\frac{C_A}{3}\right]\,{a_{\MSbar}(\mu^2)}^2\,
+\,\cdots
\label{a_bar}
\end{equation}
Recently, there have been encouraging developments \cite{papa_2_loops,Watson}
in the application of the pinch technique beyond one-loop.  This would
hopefully lead to a systematic identification of the ``skeleton effective
charge'' at higher orders, namely the determination of higher order
coefficients ($\bar{\beta}_i$ for $i\geq2$) of the $\beta$ function
$\bar{\beta}(\a)=d\a/d\ln k^2$.  This $\beta$ function should coincide
with the Gell-Mann Low function upon taking the Abelian limit $C_A =0$
(see ref.~\cite{Brodsky:1997jk}).

Being scheme invariant and free of renormalon divergence, the skeleton
expansion (\ref{skeleton_exp}) seems much favorable over the standard
perturbative QCD expansion (\ref{standard_exp}).
This advantage may become crucial in certain applications,
e.g.  for the extraction of $\alpha_s$ from event shape variables
\cite{thrust}.  However, in the absence of a concrete all-order
diagrammatic definition for the skeleton expansion in QCD,
running-coupling effects cannot be systematically resummed beyond the 
single dressed gluon level corresponding to the leading skeleton. In
particular, the momentum distribution functions of the sub-leading
skeletons are not known.
On the other hand, the BLM scale-setting procedure, which is well
defined at higher orders, can be considered as a
manifestation of the skeleton expansion. As we shall see, it is
possible in this framework to approximate the sub-leading skeleton 
terms, provided the correct skeleton scheme is used. Currently, 
since the skeleton effective charge has not been identified, the 
choice of scheme in the BLM procedure remains an additional essential 
ingredient. 

\section{BLM scale-setting}

The BLM approach \cite{BLM} is motivated by the skeleton expansion.
The basic idea is that the dressed skeleton integral (\ref{R_0}) can
be well approximated by $R_0 \simeq \a(\mu^2)\,+\,\cdots$
provided that the renormalization scale $\mu$ is properly chosen.
Indeed, by the mean value theorem \cite{Lepage:1993xa},
there exists a scale $k_0$ such that
\beq
R_0(Q^2) = \int_0^{\infty} \a (k^2) \,
\phi_0\left({k^2}/{Q^2}\right) \frac{dk^2}{k^2}
= \a(k_0^2)\int_0^{\infty} \phi_0(k^2/Q^2) \frac{dk^2}{k^2}
= \a(k_0^2)
\label{R_0_BLM}
\eeq
where the last step follows from the assumed normalization for
$\phi_i$ (\ref{norm_phi_0}).

A first approximation to $k_0$ is given by the
{\em average virtuality of the exchanged gluon},
\beq
k_{0,0}^2 = Q^2\exp\left(
\int_0^{\infty}\ln\frac{k^2}{Q^2}\phi_0(k^2/Q^2)\frac{dk^2}{k^2}
\left/\int_0^{\infty}\phi_0(k^2/Q^2)\frac{dk^2}{k^2}\right.
\right)
= Q^2 \exp\left(-r_1^{(1)}\right)
\label{k_0}
\eeq
where $r_1^{(1)}$ is the next-to-leading coefficient of
$a_R$ in the large $\beta_0$ limit (\ref{r_i^i}).
The scale (\ref{k_0}) is called the ``leading order BLM scale''.
It can be determined directly from the $N_f$ dependent part of the
next-to-leading coefficient ($r_1$) in the perturbative
series of the observable in terms of $\a(Q^2)$,
\beq
a_R(Q^2) = \a(Q^2) +r_1 {\a(Q^2)}^2+r_2 {\a(Q^2)}^3+\cdots.
\label{standard_exs_in_bar_a}
\eeq
Thanks to the linear $N_f$ dependence of $r_1$,
it can be uniquely decomposed into a term
linear in $\beta_0$, which is related to the leading skeleton, and a free term
\beq
r_1 = r_1^{(0)} + r_1^{(1)} \beta_0,
\label{r_1}
\eeq
where both $r_1^{(1)}$ and $r_1^{(0)}$ are $N_f$ independent.
After BLM scale-setting, with $k_{0,0}^2$ given by (\ref{k_0}), one has
\beq
a_R(Q^2) = \a(k_{0,0}^2) + r_1^{(0)} {\a(k_{0,0}^2)}^2 +\cdots.
\eeq
Thus, technically, the BLM scale-setting procedure amounts, at leading
order, to eliminating the $\beta_0$ dependent part from the next-to-leading
order coefficient.  Note that although the leading order BLM scale
$k_{0,0}$ of (\ref{k_0}) has a {\em precise} meaning as the average gluon
virtuality, it is just the lowest order
approximation to $k_0$ of eq.~(\ref{R_0_BLM}).  In other words,
aiming at the evaluation of the leading skeleton
term (\ref{R_0}), setting the scale as $k_{0,0}$ is just the first step.  
This approximation can be systematically improved (see eq.~(\ref{t_0}) below)
in higher orders. 

\subsection{Multi-scale BLM and skeleton expansion correspondence}

A BLM series \cite{CSR} can be written up to arbitrary high order
\beq
a_R(Q^2) = a(k_0^2) + c_1 {a(k_1^2)}^2
+ c_2 {a(k_2^2)}^3 + c_3 {a(k_3^2)}^4+ \cdots
\label{BLM_exp}
\eeq
where the $k_i^2$ are, in general, different scales proportional
to the external scale $Q^2$ (as in (\ref{k_0})) and $c_i$ are $N_f$
independent coefficients.
The intuition behind this generalization is that each skeleton term in
(\ref{skeleton_exp}) is approximated by a corresponding term in the
multi-scale BLM series: each skeleton term may have different
characteristic momenta.  This one-to-one correspondence with the
skeleton expansion requires that the coupling $a$ will be the skeleton
effective charge $a=\a$ such that
\beq
R_i(Q^2)\equiv \a(k_i^2)^{i+1}.
\label{one_to_one_BLM}
\eeq
In this case the coefficients of sub-leading terms in (\ref{BLM_exp})
should coincide with the coefficients of the sub-leading skeleton
terms, namely $c_i=s_i$.

More generally, a BLM series can be formally written in an arbitrary scheme:
then the coupling $a$ in (\ref{BLM_exp}) can be
either defined in a standard scheme
like $\overline{\rm MS}$ or, as suggested in
\cite{CSR}, be another measurable effective charge.
However, in such cases there is no direct correspondence with the
skeleton expansion (\ref{skeleton_exp}), and as a result the forthcoming 
motivation for a unique scale setting is lost.

Let us recall how the BLM scale-setting procedure is performed
beyond the next-to-leading order \cite{CSR,GruKat}, yielding
an expansion of the form (\ref{BLM_exp}).
Suppose that the perturbative expansion of $a_R(Q^2)$
in terms of $a(Q^2)$ is given by\footnote{We work now in a generic
scheme but in contrast to (\ref{standard_exp}) we start here with
the renormalization scale $\mu=Q$ thereby simplifying the formulas
that follow.  Since the scale is tuned in the BLM procedure, this
initial choice is of little significance.  The only place where the
arbitrary renormalization scale is left at the end is in the
power series for the scales-shifts, eq.~(\ref{scale_series})
below.}
\beq
a_R(Q^2)=a(Q^2) + r_1 {a(Q^2)}^2 + r_2 {a(Q^2)}^3+ r_3 {a(Q^2)}^4 +\cdots
\label{a_R_1}
\eeq
Based on the fact that $r_i$ are polynomials of order $i$ in $N_f$
and that $\beta_0$ and $\beta_1$ are linear in $N_f$,
we can write $r_1$ as in (\ref{r_1}) and
\beq
r_2=r_2^{(0)}+r_2^{(1)}r_1^{(0)}\beta_0+r_2^{(2)}\beta_0^2+r_1^{(1)}\beta_1
\label{r_2}
\eeq
where $r_i^{(j)}$ are $N_f$ independent.  The reason for the
$\beta_1$ dependent term in (\ref{r_2}) shall become clear below.
Expanding $a(k_i^2)$ in terms of $a(Q^2)$ similarly to eq.~(\ref{rescale}),
the next-to-next-to-leading order BLM series (\ref{BLM_exp}) can be written as
\beq
a_R(Q^2)=a(Q^2)+\left(c_1+ t_0\beta_0\right)a(Q^2)^2
+\left(c_2+2t_1c_1\beta_0+t_0\beta_1+t_0^2\beta_0^2\right)a(Q^2)^3.
\label{NNLO_BLM}
\eeq
Writing the scale-shifts $t_i\equiv \ln (Q^2/k_i^2)$
as a power series in the coupling
\beq
t_i\,\equiv\, t_{i,0}\,+\,t_{i,1}\,a(Q^2)+\,t_{i,2}\,a(Q^2)^2+\cdots
\label{scale_series}
\eeq
where $t_{i,0}$ are assumed to be $N_f$ independent, we get
\begin{eqnarray}
\label{NNLO_BLM2}
a_R(Q^2)&=&a(Q^2)+\left(c_1+ t_{0,0}\beta_0\right)a(Q^2)^2\\ \nonumber
&+&\left(c_2+(2t_{1,0}c_1+t_{0,1})\beta_0
+t_{0,0}\beta_1+t_{0,0}^2\beta_0^2\right)a(Q^2)^3.
\end{eqnarray}
An order by order comparison of (\ref{NNLO_BLM2}) and (\ref{a_R_1})
yields the scale shifts \hbox{$t_0=\ln(Q^2/k_0^2)$} and
\hbox{$t_1=\ln(Q^2/k_1^2)$} and the
coefficients $c_1$ and $c_2$ in terms of $r_1$ and
$r_2$ and the coefficients of the $\beta$ function of $a(Q^2)$.
The comparison at the next-to-leading order gives
\beq
c_1=r_1^{(0)}
\eeq
and
\beq
t_{0,0}=r_1^{(1)}.
\label{t_00}
\eeq
The comparison at the next-to-next-to-leading
order for the $\beta_i$ independent piece gives
\beq
c_2=r_2^{(0)}
\eeq
while for the $\beta_0$ dependent piece it gives
\beq
t_{0,1}+2t_{1,0}r_1^{(0)}+\beta_0\left(r_1^{(1)}\right)^2
=r_2^{(1)}r_1^{(0)}+\beta_0 r_2^{(2)}.
\label{NNLO_comp}
\eeq
Thanks to the explicit $\beta_1$ dependent term introduced in
(\ref{r_2}), the equality of the corresponding piece there to that
in (\ref{NNLO_BLM2}) is satisfied based on the next-to-leading order
result (\ref{t_00}).
To proceed we need to specify $t_{0,1}$ and $t_{1,0}$ such that
eq.~(\ref{NNLO_comp}) is satisfied.  Having two free parameters with
just one constraint there is apparently no unique solution.  Two natural
possibilities are the so called multi-scale BLM prescription \cite{CSR},
\begin{eqnarray}
\label{multi_scale}
t_{0,1}&=&\beta_0\left[r_2^{(2)}-\left(r_1^{(1)}\right)^2\right]\\ \nonumber
t_{1,0}&=&\frac12r_2^{(1)}
\end{eqnarray}
and the single-scale BLM prescription \cite{GruKat} where
$t_{1,0}\equiv t_{0,0}$ and
\beq
\label{single_scale}
t_{0,1}=\beta_0\left[r_2^{(2)}-\left(r_1^{(1)}\right)^2\right]-
2r_1^{(1)}r_1^{(0)}+r_2^{(1)}r_1^{(0)}.
\eeq

Having in mind the original motivation for BLM, it is interesting to
examine the case where the scheme of $a$ coincides with the skeleton
effective charge $\a$.  Then we would like to have a one-to-one
correspondence (\ref{one_to_one_BLM}) between the terms in the
BLM series (\ref{BLM_exp}) and those of the skeleton
expansion (\ref{skeleton_exp}).
The multi-scale procedure is consistent with this requirement:
the leading term $\a(k_0^2)$ in the BLM series
(\ref{BLM_exp}) represents only the leading skeleton term $R_0$ in
(\ref{skeleton_exp}), since the scale-shift
\beq
t_0=r_1^{(1)}+\left[r_2^{(2)}-\left(r_1^{(1)}\right)^2\right] \beta_0 \a(Q^2)
\label{t_0_NLO}
\eeq
involves only coefficients which are leading in the large $\beta_0$
limit and originate in $\phi_0$ (cf. eq.~(\ref{r_i^i})).
On the other hand the single-scale procedure violates this
requirement, since there $t_0$ involves (\ref{single_scale})
terms which are sub-leading in
$\beta_0$ and {\em do not belong} to the leading skeleton term $R_0$.
In fact, in order to guarantee that the scale-shift $t_0$ would
represent just the leading skeleton $R_0$ we are bound to choose
$t_{0,1}$ proportional to $\beta_0$ and thus the solution
(\ref{multi_scale}) is uniquely determined.

We see that a unique scale-setting procedure at the
next-to-next-to-leading order ($r_2$) is implied by the requirement that the
scale-shift $t_0$ should represent the leading skeleton $R_0$.
In order to continue and apply BLM at the next order ($r_3$) we
have to impose further constraints based on the structure of both $R_0$ and
$R_1$.

\subsection{BLM scale-setting for the leading skeleton}

Let us first examine the structure of the scale-shift $t_0$ by
applying BLM to a hypothetical observable that contains only an $R_0$
term of the form (\ref{R_0}).
Expanding the coupling $\a(k^2)$ under the integration sign in terms
of $a(Q^2)$ we obtain (\ref{R_0_expand}). We would like to apply BLM
to the latter series obtaining simply $\a(k_0^2)$,
with $t_0\equiv \ln(Q^2/k_0^2)=t_{0,0}+t_{0,1}\a(Q^2)+\cdots$.
Expanding $\a(k_0^2)$ we obtain from (\ref{rescale}),
\begin{eqnarray}
\label{a_k_0}
\a(k_0^2)&=&\a(Q^2)+\beta_{0} t_{0,0} \a(Q^2)^2+\left(\beta_{0} t_{0,1}
+\beta_{1} t_{0,0}+\beta_{0}^2 t_{0,0}^2\right) \a(Q^2)^3\\ \nonumber
&+&\left(\beta_{0} t_{0,2}+\beta_{1} t_{0,1}
+2 \beta_{0}^2 t_{0,0} t_{0,1}+\bar{\beta}_2
t_{0,0}+\beta_{0}^3 t_{0,0}^3+\frac52 \beta_{0} \beta_{1} t_{0,0}^2\right)
\a(Q^2)^4+\cdots
\end{eqnarray}
Comparing (\ref{R_0_expand}) with (\ref{a_k_0}) we get
\begin{eqnarray}
\label{t_0}
t_0&=&r_1^{(1)}+\left[r_2^{(2)}-\left(r_1^{(1)}\right)^2\right] \beta_0 \a(Q^2)
\\&+&\left\{
\left[r_3^{(3)}-2r_1^{(1)}r_2^{(2)}+\left(r_1^{(1)}\right)^3\right]
\beta_0^2+\frac32\left[r_2^{(2)}-\left(r_1^{(1)}\right)^2\right]\beta_1
\right\}\a(Q^2)^2+\cdots
\nonumber
\end{eqnarray}
Here we recovered the two leading orders in $t_0$ of eq.~(\ref{t_0_NLO}).
At order $\a(Q^2)^2$ we obtained an explicit dependence on both $\beta_0$
and $\beta_1$.  The combination
$r_2^{(2)}-\left(r_1^{(1)}\right)^2$ appearing at the
next-to-leading order in $t_0$ has an interpretation as the width of the
distribution $\phi_0$, assuming the latter is positive definite
(see \cite{Neu,mom}).  In general, eq.~(\ref{t_0}) can be written in
terms of central moments of the distribution $\phi_0$, defined~by
\beq
M_n\,=\,\left<\left(\ln\frac{Q^2}{k^2}-
\left<\ln\frac{Q^2}{k^2}\right>_{\phi_0}\right)^n\right>_{\phi_0}\,=\,
\left<\left(\ln\frac{k_{0,0}^2}{k^2}\right)^n\right>_{\phi_0}
\eeq
for $n\geq 2$, where
$M_1=\left<\ln\frac{Q^2}{k^2}\right>_{\phi_0}=\ln\frac{Q^2}{k_{0,0}^2}$
corresponds to
$r_1^{(1)}$ in eq.~(\ref{r_i^i}).  In terms of the central moments we
have
\begin{eqnarray}
\label{t_0_M_n}
t_0&=&M_1+M_2 \beta_0 \,\a(Q^2)+\left\{
\left[M_3+M_1M_2\right]\beta_0^2+\frac32 M_2 \beta_1
\right\}\a(Q^2)^2+\cdots \nonumber  \\
 &=& M_1 + M_2 \beta_0 \,\a(k_{0,0}^2) +
     \left\{M_3 \beta_0^2 + \frac32 M_2 \beta_1\right\} \a(k_{0,0}^2)^2+\cdots
\end{eqnarray}
where in the second step we changed the scale from $Q^2$ to the leading order
BLM scale $k_{0,0}^2$ to get simpler expressions for the coefficients of the 
$t_0$ series.
At large orders $n$ the moments $M_n$ become sensitive to extremely
large and small momenta and thus develop renormalon factorial
divergence, similarly to the standard perturbative coefficients
in eq.~(\ref{r_i^i}).  We thus see that in the BLM approach, the
scale-shift itself is an asymptotic expansion, affected by renormalons.

\subsection{BLM scale-setting for sub-leading skeletons}

Next, let us consider an $R_1$ term, given by (\ref{R_1}).
Expanding the couplings $\a(k_1^2)$ and
$\a(k_2^2)$ under the integral in terms of $\a(Q^2)$ using
(\ref{rescale}), we get (cf. the expansion of $R_0$ in eq.~(\ref{R_0_expand}))
\beq
R_1(Q^2)=\a(Q^2)^2+\beta_0 r_2^{(1)} \a(Q^2)^3+
\left(r_3^{(2)}\beta_0^2+r_2^{(1)}\beta_1)\right) \a(Q^2)^4+\cdots
\label{R_1_expand}
\eeq
where
\begin{eqnarray}
\label{r_3^2}
r_2^{(1)}&\equiv& \phi_1^{(1,0)}+\phi_1^{(0,1)} \\ \nonumber
r_3^{(2)}&\equiv& \phi_1^{(2,0)}+\phi_1^{(1,1)}+\phi_1^{(0,2)}
\end{eqnarray}
with
\beq
\phi_1^{(j,k)}\equiv \int_{0}^\infty
\left[-\ln(k_1^2/Q^2)\right]^j\,\left[-\ln(k_2^2/Q^2)\right]^k\,
\phi_1\left(k_1^2/Q^2,k_2^2/Q^2\right)
{dk_1^2\over k_1^2}\,{dk_2^2\over k_2^2}.
\eeq

The BLM scale-setting procedure can now be applied according to
(\ref{one_to_one_BLM}): $R_1(Q^2)$ given in eq.~(\ref{R_1_expand})
should be written as $\a(k_1^2)^2$.
Expanding $\a(k_1^2)^2$ in terms of $\a(Q^2)$ using (\ref{rescale})
and \hbox{$t_1= t_{1,0}+t_{1,1}\a(Q^2)+\cdots$}\, we have
\beq
\a(k_1^2)^2=\a(Q^2)^2+2t_{1,0}\beta_0\a(Q^2)^3
+\left(2t_{1,1}\beta_0+3t_{1,0}^2\beta_0^2+2t_{1,0}\beta_1\right)\a(Q^2)^4
+\cdots.
\label{a_sq}
\eeq
The comparison with (\ref{R_1_expand}) at the next-to-leading order implies
\beq
t_{1,0}=\frac{1}{2}r_2^{(1)}.
\eeq
The comparison at the next-to-next-to-leading order
then yields
\beq
2t_{1,1}\beta_0+\frac34\left(r_2^{(1)}\right)^2\beta_0^2+r_2^{(1)}\beta_1
=r_3^{(2)}\beta_0^2+r_2^{(1)}\beta_1
\eeq
which implies that $t_{1,1}$, just as $t_{0,1}$, is bound to be
proportional to $\beta_0$.  Finally we obtain the scale-shift for $R_1$,
\beq
t_1= \frac12 r_2^{(1)}+\frac12
\left[r_3^{(2)}-\frac34\left(r_2^{(1)}\right)^2\right]\beta_0\a(Q^2).
\label{t_1}
\eeq
Similarly, applying BLM to $R_2$,
\beq
R_2=\a(Q^2)^3+r_3^{(1)}\beta_0 \a(Q^2)^4+\cdots,
\label{R_2_expand}
\eeq
we get
\beq
t_2= \frac13 r_3^{(1)}.
\label{t_2}
\eeq

\subsection{Skeleton decomposition and its limitations}

Let us now return to the case of a generic observable (\ref{a_R_1})
and see that with these skeleton-expansion-correspondence constraints there
is a unique BLM scale-setting procedure.
The basic idea is that, given the existence of a skeleton expansion,
it is possible to separate the entire series into terms which originate
in specific skeleton terms.  This corresponds to a specific
decomposition of each perturbative coefficient $r_i$ similarly to (\ref{r_1})
and (\ref{r_2}).  Then the application of BLM to the separate skeleton
terms, namely representing $R_i$ by $\a(k_i^2)^{i+1}$,
immediately implies a specific BLM scale-setting procedure for the
observable.  For example, when this procedure is applied up to order
$\a(Q^2)^4$, the scale-shifts $t_i$ for $i=0,1,2$ are given by (\ref{t_0}),
(\ref{t_1}) and (\ref{t_2}), respectively.

To demonstrate this argument let us simply add up the expanded form of the
skeleton terms up to order $\a(Q^2)^4$ with $R_0$ given
by (\ref{R_0_expand}), $R_1$ by (\ref{R_1_expand}) and $R_2$ by
(\ref{R_2_expand}).  For $R_3$ we simply have at this order $R_3=\a(Q^2)^4$.
Altogether we obtain,
\begin{eqnarray}
\label{a_R_decomp}
a_R&=&\a+\left[s_1+r_1^{(1)}\beta_0\right]\a^2\\\nonumber
&+&\left[s_2+s_1r_2^{(1)}\beta_0+r_2^{(2)}\beta_0^2+r_1^{(1)}\beta_1\right]\a^3
\\ \nonumber
&+&\left[s_3+s_2r_3^{(1)}\beta_0+s_1 r_3^{(2)}\beta_0^2+r_3^{(3)}\beta_0^3
+r_1^{(1)}\bar{\beta}_2
+\frac52r_2^{(2)}\beta_1\beta_0+s_1r_2^{(1)}\beta_1\right]\a^4
\end{eqnarray}
Here we identify the notation $s_i$ which is the coefficient in front
of the skeleton term $R_i$ with $r_i^{(0)}$.  We recognize the form of $r_1$
and $r_2$ as the decompositions introduced before in eq.~(\ref{r_1})
and (\ref{r_2}) in order to facilitate the application of BLM.
We see that the skeleton expansion structure implies a specific
decomposition.  Suppose for example we know $r_1$ through $r_3$ in the
skeleton scheme.  Eq.~(\ref{a_R_decomp}) then defines a unique way to
decompose them so that each term corresponds
specifically to a given term in the skeleton expansion.  The
decomposition of $r_i$ includes a polynomial in $\beta_0$
up to order $\beta_0^i$,
\beq
s_i+\sum_{k=1}^{i} s_{i-k}r_i^{(k)}\beta_0^k
\label{deco_beta_0}
\eeq
where $s_0=1$ by the assumed normalization.
The other terms in $r_i$ in (\ref{a_R_decomp}) depend explicitly on higher
coefficients of the $\beta$ function $\bar{\beta}_j$ with $1 \leq j \leq i-1$.
Up to order $\a(Q^2)^4$ these terms depend exclusively\footnote{As we
shall see below, this is no longer true beyond this order, where the
coefficients depend on moments which appeared at previous orders, but
cannot be expressed in terms of the lower order coefficients themselves.} on
coefficients $r_j^{(k)}$ which appeared at previous orders in the
$\beta_0$ polynomials (\ref{deco_beta_0}).
Finally, we need to verify
that a decomposition of the form (\ref{a_R_decomp}) is indeed
possible.  For a generic observable $a_R$, the coefficient
$r_i$ is a polynomial of order $i$ in $N_f$.  Since the $\beta$
function coefficients $\bar{\beta}_i$ are also polynomials of maximal
order $i$, the decomposition of $r_i$ according to
(\ref{a_R_decomp}) amounts to solving $i+1$ equations with $i+1$
unknowns: $r_i^{(k)}$ with $0\leq k<i$.  Thus in general there is a
unique solution.

We see that based on the assumed skeleton structure, one can
uniquely perform a ``skeleton decomposition'' and thus also BLM scale-setting
which satisfies a one-to-one correspondence of the form (\ref{one_to_one_BLM})
with the skeleton terms.  By construction in this procedure
the scale $t_0$ is determined exclusively by the
large $\beta_0$ terms $r_i^{(i)}$ which belong to $R_0$ (see
(\ref{t_0})), $t_1$ is determined by $r_i^{(i-1)}$ terms which belong
to $R_1$ (see (\ref{t_1})), $t_2$ is determined by $r_i^{(i-2)}$
terms which belong to $R_2$, etc.

It should be stressed that {\em formally}
the decomposition (\ref{a_R_decomp}), and thus also BLM scale-setting,
can be performed in any scheme: given the
coefficients $r_i$ up to order $n$, all the coefficients
$s_i$ and $r_i^{(j)}$ for $i\leq n$ and $j\leq i$ are uniquely
determined.  No special properties of the ``skeleton
effective charge'' were necessary to show that the decomposition is
possible.  Even the assumption that for this effective charge the $\beta$
function coefficients $\bar{\beta}_i$ are polynomials of order $i$ can
be relaxed.  For example, the decomposition (\ref{a_R_decomp}) can be
formally performed in physical schemes where $\bar{\beta}_i$ are
polynomials of order $i+1$.  In this case, however, the interpretation of
$r_i^{(j)}$ in terms of the log-moments of distribution functions is
not straightforward.  It is also clear that a one-to-one
correspondence between BLM and the skeleton expansion (\ref{one_to_one_BLM})
exists only if the coupling $a$ is chosen as the skeleton
effective charge $\a$.

Let us now address several complications that limit the applicability of
the above discussion.
First, we recall the assumption we made that the entire dependence
of the perturbative coefficients on $N_f$ is related to the running coupling.
This means that any explicit $N_f$ dependence which is part of the skeleton
structure is excluded from (\ref{a_R_decomp}).  In reality there may be
skeletons with fermion loops as part of the structure, which would have
to be identified and treated separately.

Having excluded such $N_f$ dependence, we have seen that up to order
$\a(Q^2)^4$ a formal ``skeleton decomposition''~(\ref{a_R_decomp}) of the
perturbative coefficients can be performed algebraically without
further diagrammatic identification of the skeleton structure.
This is no longer true at order $\a(Q^2)^5$, where the
``skeleton decomposition'' requires 
the moments of the momentum distribution functions to be identified 
separately.  Such an
identification depends on a diagrammatic understanding of the
skeleton structure.  Looking at $R_1$,
the coefficient of $\a(Q^2)^5$ in eq.~(\ref{R_1_expand}) is
\begin{eqnarray}
\label{r_4^3}
\lefteqn{\beta_0^3\left[\phi_1^{(3,0)}+\phi_1^{(0,3)}+\phi_1^{(1,2)}
+\phi_1^{(2,1)}\right]}\\ \nonumber
&+&\beta_1\beta_0\left[2\phi_1^{(1,1)}+\frac52\left(\phi_1^{(2,0)}
+\phi_1^{(0,2)}\right)\right]
+\beta_2\left[\phi_1^{(1,0)}+\phi_1^{(0,1)}\right].
\end{eqnarray}
Writing the $\a(Q^2)^5$ term in (\ref{a_R_decomp}), one will find
as before, that the terms which depend explicitly on higher
coefficients of the $\beta$ function $\bar{\beta}_l$ with $1 \leq l \leq 3$,
contain only moments of the skeleton momentum distribution functions
$\phi_i^{(j,k)}$ which appeared in the decomposition
(\ref{a_R_decomp}) in the coefficients of $\beta_0^{j+k}\,\a^{1+i+j+k}$
at the previous orders.
However, the coefficient of $\beta_1\beta_0$ will depend on a new linear
combination of moments, different from the one identified at order
$\a(Q^2)^4$ (compare the coefficient of $\beta_1\beta_0$
in (\ref{r_4^3}) with $r_3^{(2)}$ in eq.~(\ref{r_3^2})).
Thus, strictly based on the algebraic decomposition of
the coefficients at previous orders there is no way to determine the
coefficient of $\beta_1\beta_0$ at order $\a(Q^2)^5$.  Additional information,
namely the values of $\phi_1^{(1,1)}$, $\phi_1^{(2,0)}$ and $\phi_1^{(0,2)}$
is required.  In the Abelian case, where the diagrammatic identification
of the skeleton structure is transparent, it should be straightforward to
calculate these moments separately.  In the non-Abelian theory this not
yet achievable.

\setcounter{footnote}{0}
The need to identify the skeleton structure, as a preliminary stage to writing
the decomposition of the coefficients (and thus also to BLM scale-setting)
may actually arise at lower orders {\em if several skeletons appear at
the same order}.
As mentioned in the previous section, even in the Abelian case the assumed 
form of the skeleton expansion (\ref{skeleton_exp}) is oversimplified in 
this sense and should be generalized to include several different
$s_iR_i(Q^2)$ terms at any order $i$. In the non-Abelian case one should 
expect the set of skeleton diagrams to be larger.

The simplest possibility to imagine is that the momentum distribution 
functions $\phi_i$ are $N_c$ independent. 
It is then natural to expect that at any given order there will be several 
skeletons, where each of them is characterized by its own color group 
structure.
For example, let us assume that in the case of the QCD correction 
to the photon vacuum polarization $s_1 R_1(Q^2)$ should be replaced 
by a sum of three skeleton terms, 
$s_1^{p} R_1^{p}(Q^2)+s_1^{np} R_1^{np}(Q^2)+s_1^{3g} R_1^{3g}(Q^2)$,
where the first two terms correspond to double gluon exchange (which exist in
the Abelian theory) -- the planar (p) and the non-planar (np) skeleton 
diagrams, and the last term corresponds to the three gluon vertex
skeleton diagram (which vanishes in the Abelian limit)\footnote{
Note that the three gluon vertex, which is a fundamental vertex in the theory,
cannot be considered as just renormalizing the gluon propagator and the 
quark vertex. Part of it must define a new skeleton.
This is in contrast to other diagrams appearing at this order
which just renormalize the propagators or the quark vertex, and are therefore 
not candidates for new skeleton structures.}. Each of these
three terms contributes starting at order $\bar{a}^2$. In this case, 
the skeleton decomposition of (\ref{a_R_decomp}) appears to be too
naive: each of these skeleton terms has its own momentum flow. In particular, 
if the BLM series is to have a one-to-one correspondence with the skeleton 
expansion one should write, 
\beq
a_R(Q^2) = \a(k_0^2) + s_1^p {\a(k_{1,p}^2)}^2+ s_1^{np} {\a(k_{1,{np}}^2)}^2
 + s_1^{3g} {\a(k_{1,{3g}}^2)}^2+\cdots
\label{BLM_exp_AF}
\eeq
To arrive at such a BLM series one should further decompose the
coefficients in (\ref{a_R_decomp}) as follows,
\begin{eqnarray}
\label{a_R_decomp_AF}
a_R&=&\a+\left[s_1+r_1^{(1)}\beta_0\right]\a^2 \\ \nonumber
&+& \left[s_2+\left(s_1^p r_{2,p}^{(1)}+s_1^{np} r_{2,{np}}^{(1)}
+s_1^{3g} r_{2,{3g}}^{(1)}\right)
\beta_0+r_2^{(2)}\beta_0^2+r_1^{(1)}\beta_1\right]\a^3
+\cdots
\end{eqnarray} 
where 
\begin{eqnarray}
s_1^p&\equiv&\sigma_1^{p}C_F \nonumber \\
s_1^{np}&\equiv&\sigma_1^{np}\left(C_F-\frac12 C_A\right) \nonumber \\
s_1^{3g}&\equiv&\left(\sigma_A+\frac12 \sigma_1^{np}\right)C_A
\end{eqnarray}
and
\beq
s_1=\left(\sigma_1^p+\sigma_1^{np}\right)C_F+\sigma_1^{3g}C_A
=\sigma_1^{p}C_F+\sigma_1^{np}\left(C_F-\frac12 C_A\right)+
\left(\sigma_A+\frac12 \sigma_1^{np}\right)C_A.
\eeq
Here the combination $\left(C_F-\frac12 C_A\right)$ corresponding
to the non-planar skeleton (np) is suppressed in the large $N_c$ 
limit\footnote{In ${\rm SU}(N_c)$ the combination 
$\left(C_F-\frac12 C_A\right)$ is
sub-leading in $N_c$ compared to $C_A=N_c$ and $C_F=({N_c}^2-1)/(2N_c)$.}.
Since the two Abelian parts of $s_1$, namely $s_1^p$ and $s_1^{np}$, 
are separately calculable, the coefficients $r_{2,p}^{(1)}$, 
$r_{2,np}^{(1)}$ and $r_{2,{3g}}^{(1)}$ are uniquely determined from $r_{2}$. 
Thus, in this example the color group structure plus the Abelian
skeleton decomposition allow one to determine the non-Abelian skeleton
decomposition. In the general case, where more skeleton structures are
possible, this information will not suffice, and the decomposition of the
coefficients will require a more complete understanding of the 
non-Abelian skeleton expansion.

To summarize, we have seen that by tracing the flavor dependence of the
perturbative coefficients in the skeleton scheme, one can identify
the contribution of the different skeleton terms.
This procedure allows us to ``reconstruct'' the skeleton expansion
algebraically from the calculated coefficients as summarized by
eq.~(\ref{a_R_decomp}).  This decomposition implies a unique BLM
scale-setting which has a  one-to-one correspondence with the skeleton
expansion. We also learned that there are several limitations to the
algebraic procedure which can probably be resolved only by explicit
diagrammatic identification of the skeleton structures and the skeleton
effective charge.  These limitations include the need to
\begin{description}
\item{a) } treat separately contributions from skeleton structures
which involve fermion loops (in the Abelian case these are just the
light-by-light type diagrams)
\item{b) } identify separately the different moments
$\phi_i^{(j,k)}$ of a given momentum distribution function which appear
as a sum (with any $j$ and $k$ such that $j+k=n$) in the perturbative
coefficients of $\beta_0^{n}\,\a^{1+i+n}$
\item{c) } identify separately the contributions of different skeleton
terms which happen to appear at the same order in $\a$.
\end{description}


\section{Using the ECH method in the framework of the skeleton expansion}

As we saw in the previous section, the essential ingredient of the BLM
approach, which crucially relies on the skeleton expansion, is to 
disentangle running-coupling effects and treat them separately from the 
remaining expansion. Technically, this is realized by
performing a skeleton decomposition. Running coupling effects can then 
be resummed in various ways, aiming at the approximation of the skeleton 
integrals $R_i$. First, there is the possibility to perform a full
all-order resummation by evaluating the integrals using some regularization 
in the infrared region (see e.g.~\cite{thrust}). This, however, requires 
the computation of the corresponding momentum distribution function, 
which can currently be done only at the level of the leading skeleton term. 
At the more modest level, the skeleton decomposition
itself (\ref{a_R_decomp}) provides some information about the first few 
moments of the momentum distribution functions which is then used (section~3)
to perform BLM scale-setting. Alternatively, the same information can be 
used to approximate the skeleton terms $R_i(Q^2)$ in the ECH method, 
which is particularly fit to deal with running coupling 
effects \cite{ECH} (see also \cite{Maxwell_new}). 
We shall see that this method has close relations with the 
scale setting procedure, but it also has some advantages over the latter.

In this section we demonstrate how the ECH method can be used to provide
resummation of running coupling effects in the framework of the 
skeleton expansion. The basic idea is that each skeleton term $R_i(Q^2)$ in 
our ansatz (\ref{skeleton_exp}) is a renormalization-group invariant 
effective-charge raised to some power, i.e. one writes 
$R_i(Q^2)\equiv \left(a_{R_i}(Q^2)\right)^{i+1}$ instead of
(\ref{one_to_one_BLM}). 
Thus \hbox{$a_{R_i}(Q^2)\equiv \left(R_i(Q^2)\right)^{1/(i+1)}$} can be simply 
be evaluated in the ECH method~\cite{ECH}, avoiding any explicit 
scale-setting procedure. In this method $a_{R_i}$ is computed by inverting 
the integrated renormalization-group equation,
\beq
\ln Q^2/\Lambda_{R_i}^2\,=\,\int_0^{a_{R_i}}\frac{da}{\beta_{R_i}(a)}.
\eeq
Finally, the observable $a_R(Q^2)$ will be written as
(cf. eq.~(\ref{skeleton_exp}) and (\ref{BLM_exp})):
\beq
a_R(Q^2) = a_{R_0}(Q^2) + s_1 a_{R_1}(Q^2)^2 + s_2 a_{R_2}(Q^2)^3+\cdots.
\label{ECH_exp}
\eeq

Consider first the effective charge defined by the leading
skeleton term $a_{R_0}\equiv R_0$ as expanded in (\ref{R_0_expand}). From the
next-to-leading order coefficient in this equation it follows that the  
ratio between the two scale parameters characterizing $a_{R_0}$ and
$\bar{a}$ is: 
\beq
\Lambda_{R_0}^2/\bar{\Lambda}^2=e^{-r_1^{(1)}}.
\label{R_0_scale_ratio}
\eeq
This ratio is fully determined by the center of the momentum distribution 
function (which is also the leading order BLM scale-shift $t_{0,0}$,
cf. eq.~(\ref{k_0})) and 
is not modified at higher orders. The latter affect just 
the corresponding ECH $\beta$ function, \hbox{$\beta_{R_0}(a_{R_0})\equiv 
da_{R_0}/\ln Q^2$}.
Using the next-to-next-to-leading order expansion of $a_{R_0}$ in terms of $\a$
and applying the general relation between effective charges \cite{ECH}, 
we have
\beq
\beta_2^{R_0}=\bar{\beta}_2 +\beta_0\left(r_2- r_1^2\right)-\beta_1 r_1,
\eeq
where $\bar{\beta}_2$ and $\beta_2^{R_0}$ are the three-loop $\beta$
function coefficients of the skeleton coupling and of $a_{R_0}$, respectively.
Using now $r_1$ and $r_2$ of eq.~(\ref{R_0_expand}) we obtain
\beq
\beta_2^{R_0}=\bar{\beta}_2 +
\left[r_2^{(2)}- \left(r_1^{(1)}\right)^2\right]\beta_0^3=\bar{\beta}_2
+ M_2\beta_0^3.
\label{ECH_width_relation}
\eeq
This means that for any momentum distribution $\phi_0$,
$\beta_2^{R_0}$ is simply a sum of a
universal piece $\bar{\beta}_2$, which characterizes the skeleton
coupling, and an observable-dependent piece, namely the
width of $\phi_0$ (see section~3) multiplied by $\beta_0^3$.

Recall that the three-loop $\beta$ function coefficient in the skeleton 
scheme $\bar{\beta}_2$ is a polynomial of order $2$ in
$\beta_0$, namely 
$\bar{\beta}_2=\bar{\beta}_{2,0}+\bar{\beta}_{2,1}\beta_0+\bar{\beta}_{2,2}
\beta_0^2$ (see the footnote following eq.~(\ref{r_i^i})). 
Therefore $\beta_2^{R_0}$ is given by
\begin{eqnarray}
\label{ECH_width_relation_expanded_R0}
\beta_2^{R_0}=
\bar{\beta}_{2,0}+\bar{\beta}_{2,1}\beta_0+\bar{\beta}_{2,2}\beta_0^2+
\left[r_2^{(2)}- \left(r_1^{(1)}\right)^2\right]\beta_0^3.
\end{eqnarray}
In the large $\beta_0$ limit $\beta_2^{R_0}$ is dominated by the
last term, namely by the width of momentum distribution $\phi_0$. In this 
case it is therefore the width which controls the convergence of the 
ECH $\beta$ function, i.e. the accuracy of the calculated effective charge.
Note that the same parameter controls the accuracy of the leading order BLM
approximation~\cite{Neu,mom}.
Away from the large $\beta_0$ limit, a small width implies 
proximity of $\beta_2^{R_0}$ and $\bar{\beta}_2$ (see Appendix A). 
Thus only if the universal $\bar{\beta}_2$ is not large, a small width implies 
smallness of $\beta_2^{R_0}$, i.e. good convergence of the effective 
charge approach applied to~$R_0$.
Similarly at the four-loop level, one gets
\beq
\beta_3^{R_0}=\bar{\beta}_3 +
2M_3\beta_0^4+5M_2\beta_1\beta_0^3.
\label{ECH_width_relation_4loop}
\eeq
As usual~\cite{ECH} the effective-charge $a_{R_0}$ is characterized by
the scale ratio $\Lambda_{R_0}^2/\bar{\Lambda}^2$ and $\beta$
function coefficients. The same holds for higher skeleton terms. 
For example, it follows from (\ref{R_1_expand}) that 
$a_{R_1}\equiv \left(R_1\right)^{1/2}$ is characterized by
$\Lambda_{R_1}^2/\bar{\Lambda}^2=e^{-\frac12 r_2^{(1)}}$ and 
$\beta_2^{R_1}=\bar{\beta}_2+
\frac12\left[r_3^{(2)}-\frac34\left(r_2^{(1)}\right)^2\right]\beta_0^3$. 
Note that these are the same combinations appearing in the BLM scale-shift for 
$R_1$, eq.~(\ref{t_1}).

It is also worth noting that the suggested effective charge approach 
yields a result identical to the BLM scale setting method applied 
in the skeleton scheme, in the approximation where the $\beta$ 
functions of the effective charges associated to the various skeletons 
$\beta^{R_j}$ are all replaced by the skeleton coupling $\beta$ function, 
$\bar{\beta}$. This is equivalent to assuming that, except the average, 
all the central moments of the characteristic functions vanish identically. 
Then both approaches effectively yield a multi-scale series where the scales
correspond to the average momentum flowing in each skeleton diagram,~$k_{i,0}$.

To conclude, we have shown that the explicit scale-setting procedure can be 
replaced by the ECH method.
One advantage is that the latter does not suffer from the scheme and scale
ambiguities  still present (see the footnote before eq.~(\ref{a_R_1})) 
in the series for the BLM scales. We stress that 
our example here heavily relies on the specific ansatz assumed for the 
skeleton expansion. However, contrary to the BLM 
scale setting method, the suggested effective charge approach would
apply equally well to more general cases where e.g. the two couplings in 
eq.~(\ref{R_1}) are different.


\section{BLM and conformal relations}

Let us now consider the general BLM scale-setting method,
where the scheme is not necessarily the one of the skeleton effective
charge, and no correspondence with the skeleton expansion is sought for.
Then any scale-setting procedure which yields an expansion of
the form (\ref{BLM_exp}) with
$N_f$ independent $c_i$ coefficients and scale-shifts which are power
series in the coupling (\ref{scale_series}) is legitimate.  We saw that
under these requirements there is no unique procedure for setting
the BLM scale beyond the leading order ($k_{0,0}$).
Nevertheless, as we now show, the coefficients $c_i$ are uniquely defined.
In fact, the $c_i$ have a precise physical interpretation as the
``conformal coefficients'' relating $a_R$ and $a$ in a conformal
theory defined by
\beq
\beta(a)\,=\,-\beta_0a^2\,-\,\beta_1a^3+\cdots\,=\,0.
\label{beta_eq_0}
\eeq

To go from real-world QCD to a situation where such a conformal
theory exists one has to tune $N_f$: when $N_f$ is set large
enough (but still below $\frac{11}{2} N_c$, the point where asymptotic
freedom is lost) $\beta_1$ is negative while $\beta_0$ is positive and small.
Then the perturbative $\beta$ function has a zero at
$a_{\FP}\simeq-\beta_0/\beta_1$; i.e.  there is a non-trivial
infrared fixed-point~\cite{Gross:1973ju}--\cite{super}.
The perturbative analysis is justified if $\beta_0$,
and hence $a_{\FP}$, is small enough.

Physically, the existence of an infrared fixed-point in QCD
means that correlation functions are {\em scale invariant} at large
distances.  This contradicts confinement which requires a characteristic
distance scale.  In particular, when $\beta_0\to 0$ the infrared coupling is
vanishingly small.  Then it is quite clear that a non-perturbative
phenomenon such as confinement will not persist.
The phase of the theory where the infrared physics is controlled by a
fixed-point is called the conformal window.  In this work we are not
concerned with the physics in the conformal
window\footnote{In \cite{super} this phase is investigated from
the point of view of perturbation theory in both QCD
and supersymmetric QCD.}. We shall just use formal
expansions which have a particular meaning in this phase.

The BLM coefficients $c_i$ are by definition $N_f$-independent.
Therefore the expansion of $a_R$ according to
eq.~(\ref{BLM_exp}) is valid, with the same $c_i$'s both
in the real world QCD and in the conformal window. In the conformal
window a generic
coupling $a(k^2)$ flows in the infrared to a well-defined limit
$a(k^2=0)\equiv a_{\FP}$.  In particular, eq.~(\ref{BLM_exp}) becomes
\beq
a_R^{\FP}=a_{\FP} + c_1 {a_{\FP}}^2 + c_2 {a_{\FP}}^3 + c_3
{a_{\FP}}^4 +  \cdots
\label{fp_relation}
\eeq
where we used the fact that the $k_i$'s are proportional to $Q$, which
follows from their definition $k_i^2=Q^2 \exp(-t_i),$ together with the
observation that the scale-shifts $t_i$ in (\ref{scale_series})
at any finite order are just constants when $a(Q^2)\longrightarrow
a_{\FP}$. Eq.~(\ref{fp_relation}) is simply the perturbative relation
between the fixed-point values of the two couplings
(or effective charges) $a_R$ and $a$.

Note that in this discussion we ignored the complication discussed at the
end of section~3, concerning the possibility of applying
BLM scale-setting in the case of several skeletons contributing at the
same order (cf. eq.~(\ref{BLM_exp_AF})).  In this case
the argument above holds as well, while the conformal coefficients
will be the sum of all BLM coefficients appearing at the corresponding
order.  For the example considered
in section~3, (\ref{a_R_decomp_AF}),
we would then have $c_1=s_1=s_1^p+s_1^{np}+s_1^{3g}$.

According to the general argument above, the BLM coefficients
(\ref{BLM_exp}) should coincide with the
conformal coefficients in (\ref{fp_relation}).  In the next section we
calculate conformal coefficients directly and check this statement
explicitly in the first few orders.


\section{Calculating conformal coefficients}

Let us now investigate the relation between the conformal
coefficients $c_i$ appearing in (\ref{fp_relation}) and the
perturbative coefficients $r_i$.

For this purpose, it is useful to recall the Banks-Zaks expansion:
solving the equation $\beta(a)=0$ in (\ref{beta_eq_0}) for such $N_f$
where
$\beta_0$ is small and positive and $\beta_1$ is negative, we obtain:
$a_{\FP} \simeq -\beta_0/\beta_1>0$.  If we now tune $N_f$ towards the limit
$\frac{11}{2}N_c$ from below, $\beta_0$ and therefore $a_{\FP}$ become
vanishingly small, which justifies the perturbative analysis
\cite{BZ,BZ_grunberg}.
In particular, it justifies neglecting higher orders in the $\beta$
function as a first approximation.
In order to take into account the higher orders in the $\beta$ function,
one can construct a power expansion solution of the equation
$\beta(a)=0$, with the expansion parameter as the leading order
solution,
\beq
a_0\equiv
-\frac{\beta_0}{\left.\beta_1\right\vert_{\beta_0=0}}
=\frac{\beta_0}{-\beta_{1,0}}.
\label{a_0_def1}
\eeq
In the last equality we defined
$\beta_1\equiv\beta_{1,0} + \beta_{1,1}\beta_0$ where $\beta_{i,j}$
are $N_f$-independent.  Similarly, we define\footnote{We recall that in
the skeleton scheme $\bar{\beta}_{2,3}=0$.} for later use
\beq
\beta_2\equiv\beta_{2,0} + \beta_{2,1} \beta_0 + \beta_{2,2}\beta_0^2
+ \beta_{2,3}\beta_0^3.
\label{beta2_ij}
\eeq
We shall assume that the coupling $a$ has the following Banks-Zaks expansion
\beq
a_{\FP}=a_0 \,+\, v_1 a_0^2 \,+\, v_2 a_0^3\, +\, v_3 a_0^4 + \cdots
\label{BZ1}
\eeq
where $v_i$ depend on the coefficients of $\beta(a)$, see
e.g.~\cite{FP}.  For instance, the first Banks-Zaks coefficient is
\beq
\label{v_1}
v_1=\beta_{1,1}-\frac{\beta_{2,0}}{\beta_{1,0}}.
\eeq

Suppose that the perturbative expansion of $a_R(Q^2)$
in terms of $a(Q^2)$ is given by
\beq
a_R(Q^2)=a(Q^2) + r_1 {a(Q^2)}^2 + r_2 {a(Q^2)}^3 +\cdots
\label{a_R}
\eeq
Based on the fact that $r_i$ are polynomials of order $i$ in $N_f$,
and that $a_0$ is linear in $N_f$, one can uniquely write a
decomposition of $r_i$ into polynomials in $a_0$ with
$N_f$-independent coefficients
\begin{eqnarray}
\label{dij}
r_1&=&r_{1,0} + r_{1,1} a_0 \\ \nonumber
r_2&=&r_{2,0} + r_{2,1} a_0 + r_{2,2} {a_0}^2 \\ \nonumber
r_3&=&r_{3,0} + r_{3,1} a_0 + r_{3,2} {a_0}^2 +r_{3,3} {a_0}^3
\end{eqnarray}
and so on.
For convenience we expand here in $a_0$ rather than in $\beta_0$.
The relations with the ``skeleton decomposition'' of $r_1$ and $r_2$ in
eqs.~(\ref{r_1}) and (\ref{r_2}) (or in (\ref{a_R_decomp})) are the following
\beq
\label{ur_rel_1}
\begin{array}{lll}
\begin{array}{lll}
r_{1,0}&=&r_1^{(0)}\\ \nonumber
r_{1,1}&=&-\beta_{1,0}r_1^{(1)}\\
\,
\end{array}
&\,\,\,\,\,\,\,\,\,\,\,\,\,\,\,\,\,\,\,&
\begin{array}{lll}
r_{2,0}&=&r_2^{(0)}+\beta_{1,0}r_1^{(1)}\\ \nonumber
r_{2,1}&=&-\beta_{1,0}r_2^{(1)}r_1^{(0)}
-\beta_{1,0}\beta_{1,1}r_1^{(1)}\\\nonumber
r_{2,2}&=&\beta_{1,0}^2r_2^{(2)}.
\end{array}
\end{array}
\eeq
For $r_3$ we have, based on (\ref{a_R_decomp}),
\beq
r_{3,0}=r_3^{(0)}+r_2^{(1)}r_1^{(0)}\beta_{1,0}+r_1^{(1)}\beta_{2,0}.
\label{ur_d3}
\eeq

Using eq.~(\ref{a_R}) at $Q^2=0$ with (\ref{dij}) and the Banks-Zaks
expansion for $a_{\FP}$ (\ref{BZ1}), it is straightforward to obtain
the Banks-Zaks expansion for $a_R^{\FP}$
\beq
a_R^{\FP}=a_0 \,+\, w_1 a_0^2 \,+\, w_2 a_0^3 \, +\, w_3 a_0^4 + \cdots
\label{BZ2}
\eeq
with
\begin{eqnarray}
\label{wi}
w_1&=&v_1+r_{1,0}\\ \nonumber
w_2&=&v_2+2r_{1,0}v_1 +r_{1,1} +r_{2,0} \\ \nonumber
w_3&=&v_3+2r_{1,0}v_2+r_{1,0}{v_1}^2+2r_{1,1}v_1+3r_{2,0}v_1+r_{2,1}+r_{3,0}
\end{eqnarray}

Having the two Banks-Zaks expansions, one can also construct the
series which relates two effective charges $a_R^{\FP}$ and
$a_{\FP}$ at the fixed-point.  Inverting the series in (\ref{BZ1}) one
obtains $a_0$ as a power series in $a_{\FP}$,
\beq
a_0=a_{\FP}\,+\, u_1 a_{\FP}^2 \,+\, u_2 a_{\FP}^3\, +\, u_3 a_{\FP}^4 + \cdots
\label{BZ3}
\eeq
with $u_1=-v_1$ and $u_2=v_1^2-v_2$ etc.
Substituting eq.~(\ref{BZ3}) in (\ref{BZ2}) one obtains the
``conformal expansion'' of $a_R^{\FP}$ in terms of $a_{\FP}$
according to eq.~(\ref{fp_relation}) with
\begin{eqnarray}
\label{ci}
c_1&=&r_{1,0}\\ \nonumber
c_2&=&r_{1,1} +r_{2,0} \\ \nonumber
c_3&=&-r_{1,1}v_1+r_{2,1}+r_{3,0} \\ \nonumber
c_4&=&2r_{1,1}{v_1}^2-r_{1,1}v_2-r_{2,1}v_1+r_{2,2}+r_{3,1}+r_{4,0}
\end{eqnarray}
Thus the coefficients $v_i$ of the Banks-Zaks expansion (\ref{BZ1}) and the
coefficients $r_i$ of (\ref{a_R}) are sufficient to determine the conformal
coefficients $c_i$ to any given order.

Clearly, the Banks-Zaks expansions (\ref{BZ1}) and (\ref{BZ2})
and the conformal expansion of one fixed-point in terms of another
(\ref{fp_relation}) are closely related.
Strictly speaking, both type of expansions are meaningful only in the
conformal window.
However, we saw that the coefficients of (\ref{fp_relation}) coincide with
the ones of the BLM series (\ref{BLM_exp}) which is useful in real
world QCD.  We recall that the general argument in the previous section
does not depend on the specific BLM scale-setting prescription used,
provided that the scales $k_i$ are proportional to $Q$
and the $c_i$'s are $N_f$ independent.  Comparing explicitly
$c_1$, $c_2$ and $c_3$ in eq.~(\ref{ci}) with the BLM coefficients
obtained in the previous section, namely $c_i=r_i^{(0)}$,
we indeed find that they are equal (compare using eq.~(\ref{ur_rel_1}),
(\ref{ur_d3}) and (\ref{v_1})).  In particular, the ``skeleton
decomposition'' of eq.~(\ref{a_R_decomp}), which can be formally
performed in any scheme, provides an alternative way to
{\em compute} conformal coefficients.


\section{Examples}

The skeleton expansion assumption implies that the skeleton (conformal)
coefficients $s_i$ are free of running coupling effects. In particular,
contrary to the standard perturbative coefficients in a standard
scheme such as $\overline{\rm MS}$, the large order 
behavior of conformal coefficients is not dictated by renormalon 
factorial increase, and should therefore be softer. 

In other words, the effective convergence of the fixed-point relation 
(\ref{fp_relation}) where $a$ is taken as the skeleton coupling
effective charge $\a$ is expected to be better than standard
perturbative expansions. As we shall see in section~7.3,
this expectation is not restricted to the skeleton scheme but 
applies also to general conformal relations, e.g. between two 
physical effective charges. In addition, if we assume that the skeleton
coupling $\beta$ function itself is renormalon-free, it follows that
also the Banks-Zaks expansion of a generic physical quantity $a_R$ is
renormalon-free. This is because the latter assumption implies that  
the Banks-Zaks expansion of $\a$ (\ref{BZ1}) is free of renormalons, and
then, by substituting it in the renormalon-free conformal relation between the
observable $a_R$ and $\a$, one recovers the Banks-Zaks expansion of
$a_R$, which must therefore be renormalon-free as well.

Thus, the general expectation is that all conformal and Banks-Zaks 
relations are free of renormalons and have better convergence properties. 
Our purpose here is to examine through available examples in QCD whether 
this expectation is realized. Indeed, as we recall below, it has been noted by 
several authors (e.g. in \cite{CSR,Crewther_obs,BLM_Crewther,CaSt,FP}) that
conformal coefficients and Banks-Zaks coefficients are typically small.
We would like to interpret these observations based on the assumed skeleton
expansion and relate them to the absence of renormalons.
As concrete examples we shall concentrate on the following
observables:
\begin{description}
\item{a) } The Adler D-function,
\beq
D(Q^2)=Q^2\frac{d\Pi(Q^2)}{dQ^2}
\equiv N_c \sum_fe_f^2\left[1+\frac34\,C_F\,a_D\right]
\label{D_def}
\eeq
where $a_D$ is normalized as an effective charge, and $\Pi(Q^2)$ is the
electromagnetic vacuum polarization,
\beq
4\pi^2 i\int d^4x\, e^{iq\cdot x}\left\langle 0
\vert T\left\{j^\mu(x),j^\nu(0)\right\}\vert 0 \right\rangle =
\left(q^\mu q^\nu-q^2g^{\mu\nu}\right)\,\Pi(Q^2).
\eeq
\item{b) } The polarized Bjorken sum-rule for electron nucleon
deep-inelastic scattering,
\beq
\int_0^1 \left[g_1^p(x,Q^2)-g_1^n(x,Q^2)\right]dx\equiv
\frac{g_A}{6}\left[1-\frac34\,C_F\,a_{g_1}\right].
\label{Bj_def}
\eeq
\item{c) } The non-polarized Bjorken sum-rule for
neutrino nucleon deep-inelastic scattering,
\beq
\int_0^1 dx\left[F_1^{\bar{\nu} p}(x,Q^2)-F_1^{\nu n}(x,Q^2)\right]
\equiv 1-\frac{C_F}{2}\,a_{F_1}.
\label{F1_def}
\eeq
\item{d) } The static potential,
\beq
V(Q^2)\equiv -4\pi^2 C_F \frac{a_V}{Q^2}.
\label{V_def}
\eeq
\end{description}
In all four cases perturbative calculations have been performed (refs.
\cite{Ree_NNLO} through \cite{V_NNLO}, respectively) up
to the next-to-next-to-leading order $r_2$ in eq.~(\ref{standard_exp}).

For later comparison with conformal relations, we quote some numerical
values of the coefficients in the standard perturbative expansion
in $a_{\MSbar}\equiv a_{\MSbar}(Q^2)$
for the vacuum polarization D-function (\ref{D_def})
\beq
\label{D_MSbar}
\begin{array}{lcrlcrll}
a_{D}=a_{\MSbar}&+&d_1& a_{\MSbar}^2&+ &d_2&
a_{\MSbar}^3+\cdots&\\
& & 2.0&      &&18.2&& N_f=0    \\
& & 1.6&      && 6.4&&N_f=3    \\
& & 0.14&     && -27.1&& N_f=16    \\
& & 1.06&     && 14.0&& N_f=0..16
\end{array}
\eeq
and for the polarized Bjorken sum-rule (\ref{Bj_def})
\beq
\label{Bj_MSbar}
\begin{array}{lcrlcrll}
a_{g_1}=a_{\MSbar}&+&k_1& a_{\MSbar}^2&+ &k_2&
a_{\MSbar}^3+\cdots&\\
& & 4.6&     && 41.4   && N_f=0    \\
& & 3.5&     && 20.2  && N_f=3    \\
& & -0.75&   && -34.8 && N_f=16   \\
& & 2.1&     && 21.0   && N_f=0..16
\end{array}
\eeq
where in the first three lines in (\ref{D_MSbar}) and (\ref{Bj_MSbar})
the coefficients are evaluated
at given $N_f$ values, while the last
line corresponds to an average of $\vert r_i\vert$
in the range $N_f=0$ through~$16$.

We see that the coefficients in a running coupling
expansion in the $\overline{\rm MS}$ scheme increase fast already at
the available next-to-next-to-leading order. This increase has been
discussed in connection with renormalons, for example in
\cite{Beneke}.
A priori, it is hard to expect that the large-order
behavior of the series will show up already in the first few leading orders.
We mention, however, that in ref.~\cite{PBB} the Bjorken sum rule
series (for $N_f=3$) was analyzed in the Borel plane based on the three known
coefficients, indicating that the first infrared renormalon at
$p=1$ does show up.

\subsection{The Banks-Zaks expansion}

Let us now compare the magnitude of the coefficients in the standard
expansion, e.g. in eqs.~(\ref{D_MSbar}) and~(\ref{Bj_MSbar}), 
to that of conformal coefficients. For the latter, one can choose
to examine conformal relations between effective charges (see
section~7.3) or -- the Banks-Zaks expansion.   
 
The Banks-Zaks expansion for the fixed-point value of the vacuum
polarization D-function (\ref{D_def}) is
\beq
\label{D_BZ}
\begin{array}{llllllll}
a^{\FP}_{\rm D}=& {  a_0} &+& 1.22 \,{  a_0}^{2} &+&
0.23 \,{ a_0}^{3} &+& \cdots
\end{array}
\eeq
whereas for the Bjorken sum-rule it is
\beq
\label{Bj_BZ}
\begin{array}{llllllll}
a^{\FP}_{g_1}=&{  a_0} &+& 0.22\, {  a_0}^{2} &-&
1.21 \, 
{  a_0}^{3} &+& \cdots.
\end{array}
\eeq
Comparing (\ref{Bj_BZ}) and (\ref{D_BZ}) with the corresponding
running coupling expansions in $\overline{\rm MS}$, namely
(\ref{D_MSbar}) and (\ref{Bj_MSbar}), the difference in magnitude of
the coefficients is quite remarkable~\cite{CaSt,FP}. 
Taking into account the fact that
the coefficient of $a_0^{i+1}$ contains, among other terms, a ${C_A}^i=3^i$
term, this fast apparent convergence seems rather surprising. From this
point of view, the absence of renormalons may not be considered a sufficient 
explanation.

For the non-polarized Bjorken sum-rule
defined by (\ref{F1_BZ}), the Banks-Zaks coefficients are even smaller
\beq
\begin{array}{llllllll}
a^{\FP}_{\rm F_1}={  a_0} &-& 0.45\,{  a_0}^{2} &+& 0.16\,
{a_0}^{3} &+& \cdots
\end{array}
\label{F1_BZ}
\eeq
and exhibit an impressive cancelation of numerical terms appearing
in the running coupling coefficients \cite{FP}.
The static potential shows a different behavior.
In this case the Banks-Zaks expansion \cite{FP,super}
\beq
\begin{array}{llllllll}
a^{\FP}_{\rm V}={  a_0} &-& 0.86\,{  a_0}^{2} &+& 10.99\,
{a_0}^{3} &+& \cdots
\end{array}
\label{V_BZ}
\eeq
has a significantly larger next-to-next-to-leading order
coefficient. Taking into account the numerically large color group 
factor ${C_A}^2=9$, the magnitude of this next-to-next-to-leading order
coefficient is quite reasonable.

Another physical quantity for which the Banks-Zaks coefficients are
relatively large is the critical exponent $\hat\gamma$
\cite{BZ_grunberg,CaSt,FP,super}
\beq
\hat\gamma=\frac{1}{\beta_0}\left.\frac{d\beta(a)}{da}\right|_{a=a_{\FP}}
\eeq
where
\beq
\begin{array}{llllllll}
\hat\gamma=&a_0&+&4.75\,{a_0}^2&-&8.89\, {a_0}^3&+& \cdots
\end{array}
\label{gamma_BZ}
\eeq
Since this quantity does not depend on $Q^2$, there is no direct
comparison between a running coupling expansion and the Banks-Zaks
expansion.

To conclude, we have seen that the Banks-Zaks coefficients for 
physical quantities 
typically have smaller coefficients compared to the standard running coupling
expansion. In some cases, their convergence is surprisingly good, even taking
into account the absence of running-coupling effects. 

\subsection{Conformal relations in the skeleton scheme}

Examining the Banks-Zaks expansion we found that the coefficients
are significantly smaller than standard running-coupling coefficients.
The same conclusion would follow from examining direct conformal
relations between observables. 
The coefficients of such relations (see section~7.3) are
not only small, but also exhibit a remarkable simplicity~\cite{CSR}.  Both
the smallness and the simplicity of these coefficients seem a
natural consequence of the conformal limit. The smallness, in
particular, is naturally attributed to the absence of running-coupling
effects. 

The first step in trying to substantiate this statement in the framework
of the postulated skeleton expansion is to consider the conformal
relations in the skeleton scheme. It is natural to expect that conformal
relations between observables and the skeleton coupling will be small, 
thus explaining the above observations.

To this end, let us consider now the conformal relation in the skeleton scheme
(\ref{fp_relation}) as defined by the pinch technique. 
Since the skeleton coupling $\bar{a}$
has been identified only at the one-loop level (\ref{a_bar}), our information
on the coefficients $s_i$ is quite limited: by a direct calculation
(using the next-to-leading order coefficient $r_1$ and
either (\ref{r_1}) or (\ref{ci})) we can only determine $s_1$.
For example, for the observables defined above it is
\begin{eqnarray}
\label{exm_s1}
s_1=r_1^{(0)}=\left\{\begin{array}{lclc}
-(1/4)C_A-(1/8)C_F&=&-11/12&\,\,\,\,\,\,\,\,D\\
-(1/4)C_A-(7/8)C_F&=&-23/12&\,\,\,\,\,\,\,\,g_1\\
-(1/4)C_A-(11/8)C_F&=&-31/12&\,\,\,\,\,\,\,\,F_1\\
           -C_A&=&-3&\,\,\,\,\,\,\,\,V\\
\end{array}
\right.
\end{eqnarray}
Note the absence of a $C_F$ term in the case of the static potential.
This can be understood based on the Abelian limit, where it is known that
this effective charge coincides with the skeleton coupling (there, the
Gell-Mann Low effective charge) up to light-by-light type corrections.
Therefore the momentum distribution function of the
leading skeleton term $\phi_0$ is just a $\delta$-function,
$\phi_0(k^2)=\delta(k^2)$, and in the Abelian limit there are strictly
no ($N_f$-independent) sub-leading skeleton terms.

The higher-order coefficients $s_i$, for $i\geq
2$, depend on yet unknown characteristics of the skeleton coupling
scheme. In particular, as we discuss in Appendix~A, $s_2$
depends on the skeleton $\beta$ function coefficient
$\bar{\beta}_2$. However, as can be seen
in eq.~(\ref{expl_beta_2D}) there, the dependence on this coefficient cancels
in the difference of $s_2$ between any two observables, which is
therefore calculable.

Without a diagrammatic identification of the
skeleton structure, one cannot isolate skeletons with fermion loops
attached to
three gluons, which may appear at the order considered.
Therefore we shall just treat the entire $N_f$ dependence
(excluding Abelian light-by-light diagrams) as if it appears
due to the running coupling, according to eq.~(\ref{a_R_decomp})
where $s_2$ is $N_f$ independent.
For the observables considered above we then find:
\begin{eqnarray}
\label{delta_s_2}
s_2^{g_1}-s_2^D&=&{ \frac {3}{8}} \,{ C_F}\,{ C_A}
+ { \frac {3}{4}} \,{ C_F}^{2}=2.833 \nonumber \\
s_2^{F_1}-s_2^D&=&\left[{ \frac {43}{12}}  + {\frac {85}{6}} \,\zeta_3
- { \frac {115}{6}} \,\zeta_5\right]\,{C_A}^{2} + \left[ - 34\,\zeta_3
- { \frac {75}{8}}  + { \frac {95}{2}} \,\zeta_5\right]\,
{ C_F}\,{ C_A}\nonumber \\&+& \left[{ \frac {21}{2}}  + {\frac {47}{2}}
\,\zeta_3 - 35\,\zeta_5\right]\, {C_F}^{2}=7.045 \\
s_2^V-s_2^D&=&\left[{ \frac {1}{4}} \,\pi^{2}
+ { \frac {43}{24}} - { \frac {1}{64}} \,\pi^{4}\right]\,{ C_A}^{2}
- { \frac {25}{16}} \,{C_F}\,{C_A} +
{ \frac {23}{32}} \,{C_F}^{2}=19.66 \nonumber
\end{eqnarray}

This gives some estimate of the size of $s_2$ for these observables.
The $s_2$ coefficients turn out to be larger than the Banks-Zaks coefficients 
quoted above (as well as the conformal coefficients in the relation 
between observables). 
They can even be comparable in size to the 
next-to-next-to-leading order coefficients in~$\overline{\rm MS}$. 
Thus the assumed form of the skeleton expansion does not provide a 
satisfactory explanation for the observed smallness of conformal coefficients. 
We stress again that we mistreated here the
$N_f$ dependence which is associated with the skeleton
structure, namely fermion loops attached to three gluons.
Eventually, this will have some impact on the magnitude of the ($N_f$
dependent) skeleton coefficients $s_2$, which we cannot evaluate at present.

\subsection{Direct relations between observables}

As we saw above the knowledge about the skeleton coefficients is very limited
beyond next-to-leading order. However, there
is a way to consider systematically conformal relations
avoiding the use of the skeleton scheme.
Having renormalon-free conformal expansions (\ref{fp_relation})
for two QCD observables in terms of the skeleton effective charge $\bar{a}$,
one can eliminate the latter
to obtain a {\em direct} conformal relation between the two observables.
The existence of a skeleton expansion (\ref{skeleton_exp}) for the two
observables implies that this conformal relation is free of renormalons.

Conformal coefficients of this type can be computed either from the
Banks-Zaks expansion (\ref{ci}) or in the framework of BLM, as the
coefficients in a commensurate scale relation \cite{CSR}.
The latter can be obtained by applying BLM
directly to the perturbative relation between two observable effective
charges (and so it does not require identification of the skeleton coupling).
However, it should be noted that whereas the one-to-one correspondence 
between the BLM series and the skeleton expansion specifies a 
unique scale-setting procedure when the skeleton scheme in used, the
scale-setting procedure in direct relations between observables remains
ambiguous. As explained in sections~3 and~4, the conformal coefficients
themselves are uniquely determined, independently of the particular 
way the scales are set.

In addition to being numerically small, conformal coefficients in the
direct relations between observables turn of to be simpler~\cite{CSR}, 
in terms of color group factors and numerical $\zeta_n$ terms. This
simplicity is naturally attributed to the conformal limit.

There is one example where a direct {\em all-order} conformal
relation is known -- this is the Crewther relation
relating the vacuum polarization D-function effective charge $a_D$, defined by
(\ref{D_def}), with the polarized Bjorken sum-rule effective
charge $a_{g_1}$, defined by (\ref{Bj_def}).
The Crewther relation is \cite{Crewther,Crewther_obs,BLM_Crewther}
\beq
a_{g_1}-a_D+\frac34C_F a_{g_1}a_D=-\beta(a)T(a)
\label{crewther}
\eeq
where $T(a)$ is a power series in the coupling
\beq
\label{T}
T(a)=T_1+T_2a+T_3a^2+\cdots
\eeq
and $T_i$ are polynomials in $N_f$.

If $a_D$ has a perturbative fixed-point $a_D^{\FP}$,
then it is convenient \cite{FP} to write the r.h.s.~of
(\ref{crewther}) in terms of $a_D$.
$\beta(a_D^{\FP})=0$ and so
the r.h.s.~vanishes at $a_D=a_D^{\FP}$ corresponding to the infrared
limit. Therefore $a_{g_1}$ also
freezes perturbatively, leading to the original conformal Crewther
relation
\beq
a_{g_1}^{\FP}=\frac{a_D^{\FP}}{1+\frac34 C_F a_D^{\FP}}.
\label{crewther_conformal}
\eeq
Taking $N_c=3$ we have $C_F=\frac 43$ and then the conformal coefficients
are just {\em one} to any order in perturbation theory,
\beq
a_D^{\FP}=  a_{g_1}^{\FP} + \left(a_{g_1}^{\FP}\right)^{2}
+\left(a_{g_1}^{\FP}\right)^{3} +\cdots
\label{cr_1}
\eeq
Being a geometrical series this conformal relation provides a nice
example of a perturbative relation free of renormalon divergence. In
addition, it exemplifies the simplicity of the conformal limit: here 
the conformal coefficients do not contain any non-Abelian $C_A$ terms.

As noted in \cite{Brodsky:1999gm} (see also \cite{Crewther_obs})
it is possible to write for two generic observables A and B, at two arbitrary
scales $Q_A$ and $Q_B$, the following decomposition of the perturbative
series relating the two,
\beq
a_A =C_{AB}(a_B)+\beta(a_B)T_{AB}(a_B) .
\eeq
Here $C_{AB}$ is the ``conformal part'' of the series, i.e.
\beq
\label{C_AB}
C_{AB}(a_B)=a_B+c_1a_B^2+c_2a_B^3+\cdots
\eeq
where $c_i$ are the conformal coefficients appearing in the expansion
of $a_A^{\FP}$ in terms of $a_B^{\FP}$, and $T_{AB}(a_B)$ is a
perturbative series of the form (\ref{T}).
In other words the {\em non-conformal} part of the relation between the two
observables is factorized \cite{Crewther_obs} as $\beta(a_B)T_{AB}(a_B)$.
Taking the limit $\beta \to 0$ then gives the conformal relation.
In particular, one can write such a factorized relation between an observable
effective charge and the skeleton coupling. Then the conformal
coefficients $c_i$ in (\ref{C_AB}) are the skeleton coefficients $s_i$.
Explicitly, this can be shown based on the skeleton decomposition of
the series (\ref{a_R_decomp}),
\begin{eqnarray}
a_R &=& \left[ \a+ s_1 \bar{a}^2 + s_2 \bar{a}^3 + s_3
  \bar{a}^4+\cdots  \right]+\,\,\,
  \left[ \beta_0 \bar{a}^2 + \beta_1 \bar{a}^3 + \bar{\beta}_2 \bar{a}^4
  +\cdots \right] \\
\nonumber
 &\times & \left[ r_1^{(1)} + \left(s_1 r_2^{(1)} + r_2^{(2)} \beta_0
\right) \bar{a}
    + \left( s_2 r_3^{(1)} +s_1 r_3^{(2)} \beta_0+r_3^{(3)} \beta_0^2+
\frac{3}{2} r_2^{(2)} \beta_1  \right) \bar{a}^2
  +\cdots \right].
\end{eqnarray}

Finally, we also quote the conformal relations between the vacuum
polarization D-function and the non-polarized Bjorken sum-rule (\ref{F1_def}),
\beq
a_{D}^{\FP}=a_{F_1}^{\FP}+1.67 \left(a_{F_1}^{\FP}\right)^2
+1.57\left(a_{F_1}^{\FP}\right)^3+\cdots,
\label{D_F1_con}
\eeq
as well as the static potential (\ref{V_def}),
\beq
a_{D}^{\FP}=a_{V}^{\FP}+ 2.08 \left(a_{V}^{\FP}\right)^2
-7.16\left(a_{V}^{\FP}\right)^3+\cdots.
\label{D_V_con}
\eeq
Taking into account the ${C_A}^i=3^i$ contribution to $c_i$,
these expansions all seem well-behaved.

\subsection{Expansions in $\overline{\rm MS}$}

Finally, it is interesting to return to the expansion in
$\overline{\rm MS}$ and examine the corresponding conformal relations.
Such relations turn out to have large coefficients.
For example,
\beq
a_{D}^{\FP}=a_{\MSbar}^{\FP}-0.083 \left(a_{\MSbar}^{\FP}\right)^2
-23.22\left(a_{\MSbar}^{\FP}\right)^3+\cdots
\label{D_MSbar_con}
\eeq
and
\beq
a_{g_1}^{\FP}=a_{\MSbar}^{\FP}-0.917 \left(a_{\MSbar}^{\FP}\right)^2
-22.39\left(a_{\MSbar}^{\FP}\right)^3+\cdots
\label{Bj_MSbar_con}
\eeq
have large next-to-next-to-leading order coefficients, in a striking
contrast with the conformal relation (\ref{cr_1}) between $a_{D}^{\FP}$ and
$a_{g_1}^{\FP}$.
Note that these large conformal coefficients do not provide an
explanation of the large coefficients in (\ref{D_MSbar}) and
(\ref{Bj_MSbar}). The former are by assumption independent of $N_f$, as
opposed to the latter. For small $\beta_0$
(e.g. $N_f=16$) the negative sign (and eventually also the magnitude)
of the full coefficient can presumably be attributed to the conformal
part. However, for larger values of $\beta_0$, relevant to
real world QCD, the non-conformal part clearly dominates making the
full next-to-next-to-leading order coefficients positive.

These large conformal coefficients in (\ref{D_MSbar_con}) and
(\ref{Bj_MSbar_con}) are due to an intrinsic
property of the $\overline{\rm MS}$ coupling,
since they appear already at the level of the Banks-Zaks expansion
\cite{FP,super},
\beq
\begin{array}{llllllll}
a^{\FP}_{\MSbar} =& {  a_0} &+& 1.1366\,{  a_0}^{2} &+& 23.2656\,
{  a_0}^{3} &+& \cdots.
\end{array}
\label{MSbar_BZ}
\eeq
Note that $a^{\FP}_{\MSbar}$ has, by far, a larger
next-to-next-to-leading order Banks-Zaks coefficient compared to
any known physical effective charge.

We stress that the large next-to-next-to-leading order coefficients in
(\ref{D_MSbar_con}), (\ref{Bj_MSbar_con}) and (\ref{MSbar_BZ}) are
not associated with renormalons.
The $\overline{\rm MS}$ $\beta$ function, being defined through an
ultraviolet regularization procedure, should not be sensitive to the infrared.
Therefore infrared renormalons are not expected.
It is more difficult to draw any firm conclusion concerning the 
absence of ultraviolet
renormalons. Since there seems to be no reason to assume a skeleton structure
or any other representation in the form of an integral over a running
coupling, we suspect that ultraviolet renormalons do not exist there
as well.  

To conclude, the case of conformal relations in $\overline{\rm MS}$
teaches us not to associate automatically any large coefficient in QCD 
with running-coupling effects. Indeed, in field theory there are other 
sources of large coefficients, such as multiplicity of diagrams.


\section{Conclusions}

The fast growth of perturbative coefficients and
the related renormalization scale and scheme ambiguities of
perturbative expansions have greatly limited the predictive
power of QCD. In many cases, this divergent behavior is predominantly
due to running-coupling effects. The existence of an Abelian-like skeleton
expansion in QCD would make it possible to disentangle in a
unique way such effects, separating them from the conformal part of
the perturbative expansion of a generic physical quantity.
The effect of the running coupling could then be treated systematically
to all orders in perturbation theory in a renormalization-scheme invariant
manner by renormalon-type integrals. The normalization 
of these skeleton integrals is controlled by conformal coefficients that 
are hopefully better behaved, making the truncated skeleton expansion 
a better approximation to the physical observable compared to the standard 
perturbative expansion of the same order.

Resummation of running coupling effects has in many cases 
a significant role in phenomenology~\cite{Beneke}. Direct resummation
is currently restricted to the level of a single dressed gluon, where
the Abelian large $N_f$ limit can be used. 
The formulation of perturbation theory in the form of a skeleton
expansion has implications which go beyond the perturbative
level. In particular, it provides a natural framework to deal 
together with the resummation and the related power-corrections. 
The renormalon integral contains essential information on the type 
of power-corrections one should expect for a given observable. Moreover,
it can be used to combine~\cite{Grunberg-pow,thrust} such power corrections 
with the perturbative 
expansion avoiding double-counting or dependence on the particular 
prescription used to regularize infrared renormalons. These aspects 
were discussed in detail in~\cite{thrust} for the example of the 
average thrust.

In this paper we have concentrated on the conformal part of the
perturbative expansion, based on a postulated ansatz for the skeleton
expansion. We have shown that the ($N_f$-independent) coefficients of this
expansion and of the related BLM series have a precise
interpretation when a perturbative infrared fixed-point is 
present: they are the conformal coefficients in the series 
relating the fixed-point 
value of the observable under consideration with that of the 
skeleton effective charge. The perturbative infrared fixed-point
appearing in multi-flavor QCD allows one to calculate these conformal
coefficients through the Banks-Zaks expansion.
We stress that the identification of the skeleton coefficients with the
ones of the conformal relations defined in the small $\beta_0$ limit 
strongly relies on the particular ansatz we have taken, namely that the 
entire $N_f$ dependence originates in the running-coupling itself, 
leaving the conformal coefficients $N_f$ independent. On the other hand, the
identification of the BLM coefficients with those of the 
conformal relations of the small $\beta_0$ limit does not rely
on any additional assumption, and it holds independently of the
particular way BLM scale-setting is performed. 

Existence of an underlying skeleton structure implies that 
BLM (conformal) coefficients do not diverge factorially due
to renormalons. Of course, there can be other effects which could make
these coefficients diverge such as combinatorial factors related to
the multiplicity of diagrams. Since in QCD this type of divergence is
much softer than that of renormalons, we expect the
BLM and possibly also the Banks-Zaks expansions to
be ``better behaved''. This expectation is
supported to some extent by previous observations concerning
the smallness of the first few known BLM coefficients \cite{CSR}
and the Banks-Zaks coefficients \cite{CaSt,FP,super}. 
On the other hand, the absence of renormalons does not always seem 
to be a sufficient explanation of the observed difference between conformal and
non-conformal coefficients. At the same time, large coefficients which
are not associated with running-coupling effects do appear in QCD, e.g. in
conformal relations with the $\overline{\rm MS}$ coupling.  

The uniqueness of the skeleton coupling in QED, which is identified
as the Gell-Mann Low effective charge, is an essential ingredient of the
dressed skeleton expansion.
It is still an open question whether an Abelian-like 
skeleton expansion exists in QCD and what the constraints are 
which would determine the skeleton coupling uniquely. 
The pinch technique may provide the answer 
\cite{pinch,Watson,papa_2_loops} once it is systematically 
carried out to higher orders.
We recall that the skeleton coupling is not constrained from the 
considerations raised in this paper: the only requirement 
following from the large $N_f$ limit is that $\bar{\beta}_i$ in 
this scheme does not contain an $N_f^{i+1}$ term. 
Since the decomposition of the coefficients (\ref{a_R_decomp}) can 
be performed in any scheme yielding the moments $r_i^{(j)}$ to arbitrary 
high order, the corresponding functions
$\phi_i$ can be formally constructed, up to the limitations discussed
in section~3.4. It thus seems that one can formally associate a 
``skeleton expansion'' to any given coupling. The absence of renormalons in 
the conformal coefficients in a specific scheme implies that there are
other schemes which share the same property: it is straightforward 
to see from the definition of the skeleton terms $R_i$
that an $N_f$-independent re-scaling of the argument of the coupling
leaves the  conformal coefficients unchanged. More generally, any
``renormalon-free''  transformation of the skeleton coupling would leave
the ``skeleton coefficients'' free of renormalons. It is certainly 
interesting to find further constraints on the identity of the skeleton
effective charge in QCD.

The BLM method provides a pragmatic way to deal with running-coupling
effects beyond the single dressed gluon level. By decomposing the
perturbative coefficients in the specific way implied by
the skeleton expansion the contributions from the different
skeleton integrals as well as the conformal coefficients
can be identified. The BLM scales are then set such that there
is a one-to-one correspondence between the terms in
the BLM series and the skeletons, provided that BLM scale setting is
performed in the skeleton scheme.
As an alternative to the BLM scale setting procedure we saw that the
skeleton integrals can also be approximated by applying
the method of effective charges to the separate skeleton terms.

In practice, BLM scale-setting can also be applied in physical schemes 
yielding a commensurate scale relation. This way conformal relations, 
which have a natural, maximally convergent, form (like the conformal 
Crewther relation) can be used as a template for real-world QCD 
predictions, even if the underlying skeleton structure is not 
completely understood. Still, since the conformal relation between 
the fixed-point value of a generic
observable and that of the skeleton effective charge is
renormalon-free, it follows, upon eliminating the 
skeleton effective charge, that the coefficients in commensurate 
scale relations between observables are also renormalon-free. 
When such a conformal template is used for real-world QCD calculations, 
the effect of the non-zero $\beta-$function is to modify the 
values of the scales $k^2$ of the effective charge at 
each order of the expansion.
We stress, however, that having no more correspondence with the 
skeleton expansion, the motivation for a particular scale-setting 
procedure is lost. 

The BLM procedure cannot replace an eventual diagrammatic 
formulation of the skeleton expansion. We saw that the scale-setting
prescription depends on the ansatz for the skeleton expansion,
and any unknown concerning the form of the latter would have some impact on
the former. 
We have considered several ways in which the simple ansatz we 
introduced~(\ref{skeleton_exp})
may be generalized. This includes in particular the possibility that 
several skeleton diagrams will appear at the same order and that
certain skeleton diagrams will contain some fermion loops as part of 
their structure, making the corresponding conformal coefficients $N_f$
dependent. In addition, we have seen that even in the case of a simple 
form of the skeleton expansion, the skeleton decomposition of the 
coefficients, cannot be performed up to arbitrarily high order 
just based on the $N_f$ dependence, but rather requires some additional 
knowledge based on an explicit diagrammatic formulation. 
One should also be aware of the possibility that an Abelian-like 
skeleton expansion with a single effective charge might fail to exist in
QCD. The non-Abelian skeleton expansion may then be based on 
several dressed Green functions, namely several different effective
charges. Even in this more complicated case the most important
properties of the skeleton expansion assumed here may hold.
This includes the possibility to associate 
running-coupling effects to the various skeleton terms in a 
renormalization-group invariant way, and the
interpretation of the skeleton coefficients as conformal coefficients
when a perturbative infrared fixed-point is present. 

\vspace{50pt}

\noindent
{\large {\bf Acknowledgments}}

\vspace{15pt}

\noindent
J.R. would like to thank the SLAC theory group for
its generous hospitality during his visit last year.

\vspace{50pt}

\section*{Appendix A -- The skeleton expansion and the 
effective charge approach}

A priori, the skeleton expansion approach,
which relies on the assumption of a universal skeleton coupling, seems
antagonist to the original effective charge approach \cite{ECH} which 
treats all effective charges independently and in a symmetric manner. 
In section~4 we saw how the ECH method can be used in the framework of the 
skeleton expansion to approximate separately each skeleton term $R_i$.
Here we revisit the original ECH approach which 
attempts to evaluate the entire observable directly, and examine it from the 
point of view of the assumed skeleton expansion.

We begin by comparing the original ECH approach to the application of the 
ECH method for the leading skeleton term $R_0$. The first difference is, 
of course, in the ECH scale parameter. To facilitate the 
comparison, suppose that we start with a perturbative 
expansion~(\ref{standard_exp}) of the 
observable $a_R$ in terms of $\bar{a}(Q^2)$, with the corresponding 
coefficients $r_i$. In the original ECH approach
this implies a scale ratio of 
$\Lambda_{R}^2/\bar{\Lambda}^2=e^{-r_1/\beta_0}$. 
This can be compared with (\ref{R_0_scale_ratio}). The difference between 
the two is due to the $r_1^{(0)}$ component in the next-to-leading coefficient 
$r_1$, the component which is not associated with the leading-skeleton. 
In practice, in many cases in QCD the running-coupling component dominates 
the next-to-leading coefficient. In such cases the two scales are close.

Next, also the $\beta$ function of the ECH method,  
\hbox{$\beta_{R}(a_{R})\equiv da_{R}/\ln Q^2$} is different (beyond the 
universal two-loop order) from that of $R_0$.
At the three-loop level the latter is given in (\ref{ECH_width_relation}) 
whereas the former is
\begin{eqnarray}
\beta_2^{R}=
\beta_{2,0}^R+\beta_{2,1}^R\beta_0+\beta_{2,2}^R \beta_0^2+
\left[r_2^{(2)}- \left(r_1^{(1)}\right)^2\right]\beta_0^3,
\end{eqnarray}
where we exhibited the fact that the term
leading in $\beta_0$ is the same in $\beta_2^{R_0}$ and $\beta_2^{R}$.
As noted in section~4, in the large
$\beta_0$ limit $\beta_2^{R_0}$ is proportional to the width of the
distribution $\phi_0$, namely to 
$\left[r_2^{(2)}- \left(r_1^{(1)}\right)^2\right]$.
This remains correct also for
the $\beta$ function of the full effective charge $a_R$ 
since adding sub-leading skeleton terms would not
modify the leading ${\cal O}\left( \beta_0^3 \right)$ term.
For the four examples considered in section~7, this parameter
is given in table~\ref{tab1}.
\begin{table}[ht]
\[
\begin{array}{|c|c|c|c|c|}
\hline
\,\bar{\beta}_{2,3}\,&\,\beta_{2,3}^D
& \,\beta_{2,3}^{g_1}\,&\,\beta_{2,3}^{F_1}\,&\,\beta_{2,3}^{V}\,
\\
\hline
\hline
\,\,0 \,\,& \,\, 2.625\,\,  &\,\,2.389\,\, &\,\, 1.500\,\,  &\,\, 0\,\, \\
\hline
\end{array}
\]
\caption{Comparison of effective charge $\beta$ function coefficients
in the large $\beta_0$ approximation given by the width of $\phi_0$,
${\beta}_{2,3}=r_2^{(2)}- \left(r_1^{(1)}\right)^2$.}
\label{tab1}
\end{table}

It is natural now to consider the possibility that $R_0$ is a good
approximation to the observable $a_R$.
In the effective charge approach at the next-to-next-to-leading order,
this can be realized if $\beta_2^{R_0}$ is a good approximation to
$\beta_2^R$. In the large $\beta_0$ limit the two are equal.
Beyond the large $\beta_0$ limit one can ask whether
\beq
\beta_{2,0}^R+\beta_{2,1}^R\beta_0+\beta_{2,2}^R
\beta_0^2\,\,\,\simeq\,\,\,\bar{\beta}_2\equiv
\bar{\beta}_{2,0}+\bar{\beta}_{2,1}\beta_0+\bar{\beta}_{2,2}\beta_0^2,
\label{sum_ECH_skel}
\eeq
namely whether $\beta_{2,0}^R+\beta_{2,1}^R\beta_0+\beta_{2,2}^R
\beta_0^2$ for a
generic observable which admits a skeleton expansion is
approximately universal and close to the three-loop skeleton
coupling $\beta$ function coefficient $\bar{\beta}_2$. If this holds
for arbitrary $\beta_0$ then
\beq
\beta_{2,i}^R\simeq\bar{\beta}_{2,i}
\label{term_ECH_skel}
\eeq
for $i=0,1,2$. The violation of the equalities in (\ref{sum_ECH_skel})
and (\ref{term_ECH_skel}) is, of course, due to sub-leading terms in
the skeleton expansion $R_1$ and $R_2$. This can be seen explicitly
by substituting $r_i$ of eq.~(\ref{a_R_decomp}) in the general relation
\beq
\beta_2^{R}=\bar{\beta}_2 +\beta_0\left(r_2- r_1^2\right)-\beta_1 r_1
\eeq
to obtain the ``skeleton decomposition'' of $\beta_2^{R}$,
\beq
\beta_2^{R}=\bar{\beta}_2
+\left(s_2- s_1^2\right)\beta_0
+s_1\left(r_2^{(1)}-2 r_1^{(1)}\right)\beta_0^2
+\left[r_2^{(2)}- \left(r_1^{(1)}\right)^2\right]\beta_0^3-s_1\beta_1.
\eeq
Finally,
decomposing $\bar{\beta}_{2}$ and $\beta_1$ in terms of $\beta_0$,
we obtain\footnote{The scheme of the skeleton coupling can be parameterized
at the three-loop order \cite{ECH} by the next-to-leading order coefficient
($s_1$ and $r_1^{(1)}$) and by $\bar{\beta}_{2}$ i.e.
$\bar{\beta}_{2,i}$ for $i=0,1,2$.
Eq.~(\ref{expl_beta_2D}) then shows explicitly that
the effective charge $\beta$ function coefficient $\beta_2^R$
determines uniquely the remaining coefficients of the
``skeleton decomposition''
(\ref{a_R_decomp}) namely, $s_2$, $r_2^{(1)}$ and $r_2^{(2)}$. This
reflects the observation in section~3 that formally, the ``skeleton
decomposition'' can be performed in any scheme.}
\begin{eqnarray}
\label{expl_beta_2D}
\beta_2^{R}&=& \left[\bar{\beta}_{2,0}-\beta_{1,0}s_1\right]
+\left[\bar{\beta}_{2,1}- \beta_{1,1}s_1
+\left(s_2-s_1^2\right)\right]\beta_0\nonumber \\
&+&\left[\bar{\beta}_{2,2}
+ \left(r_2^{(1)}-2r_1^{(1)}\right)s_1
    \right]\beta_0^2
+\left[r_2^{(2)}- \left(r_1^{(1)}\right)^2\right]\beta_0^3.
\end{eqnarray}
Clearly, if for a given observable the skeleton coefficients determining the
normalization of the sub-leading skeleton terms ($s_i$) are small, then
even away from the large
$\beta_0$ limit $\beta_2^R$ will be close to $\beta_2^{R_0}$.

In order to check (\ref{term_ECH_skel}) explicitly for a given
observable, one needs to
calculate the $\beta$ function coefficients of both the observable
effective charge $\beta_{2,i}^R$ and the skeleton effective charge
$\bar{\beta}_{2,i}$. For the latter we currently know
only $\bar{\beta}_{2,0}$ (see below) and so the examination of
(\ref{term_ECH_skel}) for  $\bar{\beta}_{2,1}$ and $\bar{\beta}_{2,2}$
cannot yet be accomplished.

To obtain $\bar{\beta}_{2,0}$ we can use the
general result \cite{beta_20} or, alternatively use eq.~(\ref{expl_beta_2D}),
which is valid for a generic effective charge which admits a skeleton
expansion. The latter yields,
\beq
\bar{\beta}_{2,0}=\beta_{2,0}^{R}+\beta_{1,0}s_1.
\eeq
Using this relation for various effective charges, e.g. the
vacuum polarization D-function (\ref{D_def}) or the Bjorken sum-rule
(\ref{Bj_def}), in the skeleton coupling scheme (\ref{a_bar})
defined through the pinch technique, we obtain
\beq
\bar{\beta}_{2,0}=\frac{C_A}{512}\left(44C_F^2-88C_AC_F-301C_A^2\right),
\label{beta_20_bar}
\eeq
and for $N_c=3$,
\beq
\bar{\beta}_{2,0}=-\frac{26845}{1536}\simeq -17.477.
\eeq

Finally we check to what extent the suggested universality of the
effective charge $\beta$ function
coefficients (\ref{term_ECH_skel}) holds for the four effective
charges examined in section~6, namely the effective charges
related to the vacuum polarization D-function (\ref{D_def}) and the
polarized (\ref{Bj_def}) and non-polarized (\ref{F1_def}) Bjorken sum-rules,
as well as the static potential.
The known coefficients are listed in the following table.
\begin{table}[h]
\[
\begin{array}{|c|c|c|c|c|c|}
\hline
i\,&\,\bar{\beta}_{2,i}\,&\,\beta_{2,i}^D
& \,\beta_{2,i}^{g_1}\,&\,\beta_{2,i}^{F_1}\,&\,\beta_{2,i}^{V}\,
\\
\hline
\hline
0 &-17.477 &  -23.607 &-30.294 &  -34.753 & -37.54 \\
\hline
1 &? &  -16.032 &-11.282  &  - 6.903 & 5.366 \\
\hline
2 &? & 8.210 &8.057 &  8.783   & 11.740  \\
\hline
\end{array}
\]
\caption{Comparison of effective charge $\beta$ function coefficients.}
\label{tab}
\end{table}

Although the coefficients $\beta_{2,i}^{R}$ for these observables
have some common trend (e.g. for a given $i$ the signs are the same,
with the exception of $\beta_{2,i}^{V}$ for $i=1$) it turns out that
the fluctuations in their magnitude are rather large.
In particular, in case of $\beta_{2,0}^{R}$ for which we know the value
of the universal piece characterizing the skeleton coupling
$\bar{\beta}_{2,0}$, the latter and the contribution of the sub-leading
skeleton $R_1$ (through $s_1$ in eq.~(\ref{expl_beta_2D})) are
of the same order of magnitude. The fluctuations between different
observables are moderate only for $\beta_{2,2}^{R}$.

In~\cite{FP} it has been observed that $\beta_2^{R}$ for the
observables considered above (the static potential excluded)
exhibit very close numerical proximity, especially for $N_f=0$ through $7$.
The extent to which universality of the sort examined here
(\ref{term_ECH_skel}) holds is not enough to explain this finding of~\cite{FP}.

The proximity of $\beta_{2,2}^{R}$ for the various effective charges
implies that applying multi-scale BLM scale-setting for one observable
in terms of another, the second scale-shift $t_{1,0}$ would be close
to the leading skeleton scale-shift $t_{0,0}$. In this case the single
scale setting procedure
\cite{GruKat,BLM_Crewther} could give similar results. The same holds in the
skeleton scheme, if $\bar{\beta}_{2,2}$ is close to  $\beta_{2,2}^{R}$.
This can be deduced from eq.~(\ref{expl_beta_2D}) which gives,
\beq
\beta_{2,2}^{R}-\bar{\beta}_{2,2}
=s_1 \left(r_2^{(1)}-2r_1^{(1)}\right)=2s_1
\left(t_{1,0}-t_{0,0}\right),
\eeq
where in the last step we used the leading order results for the
scale-shifts in eq.~(\ref{t_00}) and (\ref{multi_scale}).
In this respect it is interesting to note that
applying multi-scale BLM in $\overline{\rm MS}$,
one in general obtains large values for the $t_{1,0}$ scale-shift
since $\beta_{2,2}^{\MSbar}=3.385$ is not close to $\beta_{2,2}$
of the physical effective charges.
For example, when applying BLM to $a_D(a_{\MSbar})$
one obtains $k_{0,0} = 0.707 Q$ and $k_{1,0} = 0.366\,\, 10^{-6} Q$. 
This can be contrasted, for instance, with the BLM scales for $a_D(a_V)$:
$k_{0,0} = 1.628 Q$ and $k_{1,0} = 2.487 Q$.


\end{document}